# Spin state crossover in $Co_3BO_5$

N.V. Kazak, M.S. Platunov, Yu.V. Knyazev, M.S. Molokeev, M.V. Gorev, S.G. Ovchinnikov
*Kirensky Institute of Physics, Federal Research Center KSC SB RAS, 660036 Krasnoyarsk, Russia*

Z.V. Pchelkina[1,2], V.V. Gapontsev[2], S.V. Streltsov[1,2]
[1] *M.N. Miheev Institute of Metal Physics UB RAS, 620137 Ekaterinburg, Russia*
[2] *Ural Federal University, 620002 Ekaterinburg, Russia*

J. Bartolomé[3], A. Arauzo[3,4]
[3] *Instituto de Nanociencia y Materiales de Aragón (INMA), CSIC-Universidad de Zaragoza and Departamento de Física de la Materia Condensada, 50009 Zaragoza, Spain*
[4] *Servicio de Medidas Físicas, Universidad de Zaragoza, Zaragoza, Spain*

V.V. Yumashev
*Institute of Chemistry and Chemical Technology, Federal Research Center KSC SB RAS, 660036, Krasnoyarsk, Russia*

S.Yu. Gavrilkin
*P.N. Lebedev Physical Institute of RAS, 119991 Moscow, Russia*

F. Wilhelm, A. Rogalev
*ESRF-The European Synchrotron, 71 Avenue des Martyrs CS40220, F-38043 Grenoble Cedex 9, France*

**Abstract** We have investigated the magnetic contribution of the $Co^{3+}$ ions in $Co_3BO_5$ using the X-ray magnetic circular dichroism (XMCD) and *dc* magnetic susceptibility measurements. The XMCD experiments have been performed at Co *K*-edge in $Co_3BO_5$ and $Co_2FeBO_5$ single crystals in the fully ferrimagnetically ordered phase. The Co (*K*-edge) XMCD signal is found to be related to the $Co^{2+}$ magnetic sublattices in both compounds providing strong experimental support for the low-spin $Co^{3+}$ scenario. The paramagnetic susceptibility is highly anisotropic. An estimation of the effective magnetic moment in the temperature range 100-250 K correlates well with two $Co^{2+}$ ions in the high-spin state and some orbital contribution. The crystal structure of $Co_3BO_5$ single crystal has been solved in detail at the *T* range 296-703 K. The unit cell parameters and volume show anomalies at 500 and 700 K. The octahedral environment and oxidation state of Co4 site strongly change with heating. The GGA+U calculations have revealed that at low-temperatures the system is insulating with the band gap of 1.4 eV and the $Co^{2+}$ ions are in the high-spin state, while $Co^{3+}$ are in the low-spin state. At high temperatures (T>700 K) the charge ordering disappears, and the system becomes metallic with all Co ions in $3d^7$ electronic configuration and high-spin state.

PACS number(s): 75.50.Gg, 75.30.Wx, 75.30.Gw

## 1. INTRODUCTION

The cobalt oxides belong to a large class of strongly correlated compounds showing the complex interplay between spin, charge, lattice, and orbital degrees of freedom. Like other 3d transition metals cobalt exhibits several possible oxidation states – $Co^{2+}(d^7)$, $Co^{3+}(d^6)$ and $Co^{4+}(d^5)$. A property which makes the cobalt oxides very peculiar is the ability of $Co^{3+}$ ions to accommodate various spin states, that is, low spin (LS), high spin (HS) and intermediate spin (IS). The spin state of $Co^{3+}$ appears to be very sensitive to changes in the Co-O bond length and Co-O-Co bond angle and a spin state transition could happens under action of temperature, magnetic field or external pressure. This makes the physics of the cobalt oxides very complicated being a subject of a long-standing controversy [1]. Recently, cobalt-containing oxyborates with a general formula of $M^{2+}_2M'^{3+}BO_5$ (M, M' = Co, and 3d metal ions, as well as Al, Ga, and Mg) have attracted much attention due to the discovery of their intriguing magnetic and electronic behaviors [2-7]. These materials crystallize in orthorhombic structure (Sp.gr. *Pbam*) and are isostructural to ludwigite mineral. The $M^{2+}$ and $M'^{3+}$ ions are located at the centers of edge sharing oxygen octahedra forming linear chains propagating along the short crystallographic direction ($c\approx3$ Å). The boron atoms have a trigonal-planar coordination $BO_3$, normally linked via common corners with oxygen octahedra. The cations occupy four crystallographically distinct metal sites 2*a*, 2*b*, 4*g*, and 4*h*, which are usually numbered as M1, M2, M3, and M4, respectively (Fig. 1). The first three are occupied by divalent metal ions, whereas the latter site located in the spaces between $BO_3$-groups is occupied by the trivalent cations. As a result, a dense crystal structure is formed by alternating layers of divalent and trivalent cations. The triads 3-1-3 and 4-2-4 with longest and shortest interionic distances are structurally, magnetically and electronically singled out. The common ludwigite structure allows various types of magnetic interactions involving the metal ions, including superexchange interactions, direct exchange interactions, and dipole-dipole interactions, among which the former is dominant.

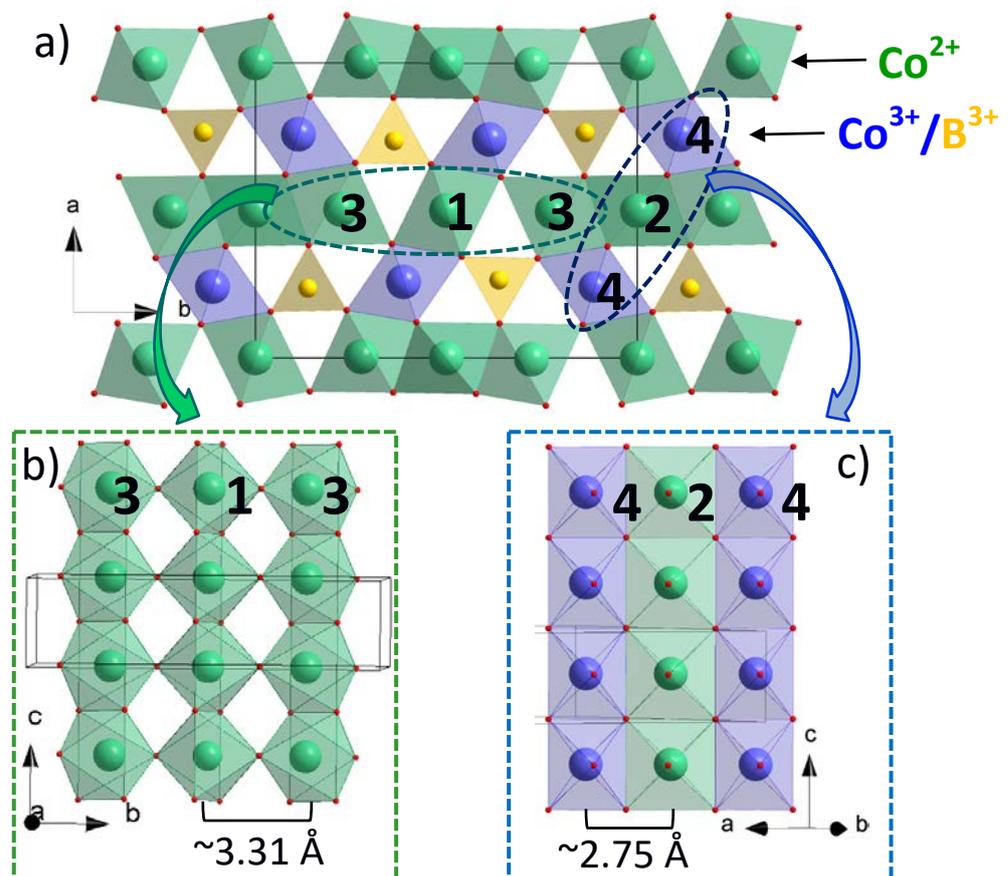

Fig. 1.a) The *ab* projection of crystal structure of $Co_3BO_5$. The green and blue spheres are octahedrally coordinated divalent and trivalent ions, respectively, that occupy different crystallographic sites 1, 2, 3, and 4. The triangle $BO_3$-groups are shown by yellow. b) The triads 3-1-3 with the longest (b) and 4-2-4 with shortest Co-Co distances (c) interionic distances are highlighted by dotted lines

The homometallic ludwigites, $Fe_3BO_5$ and $Co_3BO_5$, are the most thoroughly studied systems due to their quite distinct structural, magnetic and electronic properties [2,8]. $Fe_3BO_5$ demonstrates extremely rich physics and undergoes several transitions upon cooling from room temperature. First, a structural orthorhombic – orthorhombic transition takes place at $T_{st}$ = 283 K (*Pbam*(№55) – *Pbnm*(№62)), which is accompanied by the formation of the pairs ("dimers") of $Fe^{2+}$ and $Fe^{3+}$ ions in 4-2-4 triad [9,10]. The phase transition in $Fe_3BO_5$ detected by the X-ray diffraction also manifests itself in the temperature dependence of electrical-resistivity and hyperfine parameters, revealing well-defined anomalies at $T_{st}$ [9,11,12]. With decreasing temperature a cascade of magnetic transitions is observed by means of magnetometry, Mössbauer, and temperature-dependent neutron diffraction studies [13-15]. At $T_{N1}$=112 K the 4-2-4 spin ladder is antiferromagnetically ordered. At $T_{N2}$=74 K the ferrimagnetic ordering in the 3-1-3 spin-ladder appears. And finally, the transition to the antiferromagnetic ground state at $T_{N3}$=30 K with zero magnetic moment per unit cell is observed. The bulk magnetization measurements revealed that the anisotropy axis changes from the *a* to the *b* axis in the low-temperature antiferromagnetic phase.

On the contrary, cobalt ludwigite $Co_3BO_5$ demonstrates more conventional behavior, with the only ferrimagnetic transition at $T_N$=42 K and no structural transformations [4, 16,17]. The high magnetic uniaxial anisotropy with the *b*-axis as an easy magnetization direction was detected in the entire temperature range [15,18]. The remanent magnetization per Co atom is ~ 1.1 $\mu_B$. A small slope of magnetization at high magnetic fields seems to indicate the existence of a more complex magnetic structure.

Over the last decade, the attempts to understand the observed difference in magnetic and electronic properties of two homometalic ludwigites led to a synthesis of several new compounds: $CoMgGaBO_5$ [19], $Co_{2.4}Ga_{0.6}BO_5$ [20], $Co_{2.88}Cu_{0.12}BO_5$ [21], $Co_{4.76}Al_{1.24}BO_5$ [22], $Co_2AlBO_5$ [23], $Co_{3-x}Fe_xBO_5$ (0.0<*x*<1.0) [15,17,18,24,25], $Co_{1.7}Mn_{1.3}BO_5$ [26], $Co_5TiB_2O_{10}$ [27], and $Co_5SnB_2O_{10}$ [28] are some of them. (Here we do not mention ludwigites based on other 3*d* metal, the studies of which are also numerous). The main conclusions that can be drawn from these studies are the following: i) isovalent substitution of $Co^{3+}$ ions by nonmagnetic ions like Ga and Al or magnetic Mn causes an onset of short-range or long-range orderings, but their critical temperatures are found in the vicinity of $T_N$ characteristic of $Co_3BO_5$. The reason, apparently, is the sensitivity of the magnetic subsystem to the cation distribution. ii) The nonmagnetic substitution of type $2 \cdot Co^{3+} \rightarrow (Co^{2+}+M^{4+})$, where $M^{4+}$=Ti or Sn leads to an antiferromagnetic spin ordering at $T_N$=82 K, i.e. twice of $T_N$ of $Co_3BO_5$. Taking into account that tetravalent substitution generates $Co^{2+}$ ions at the M4 site, an increase in the Neel temperature should be attributed to the enhancement of exchange interactions via this site. iii) An unexpected effect was discovered at the substitution of $Co^{3+}$ by $Fe^{3+}$ ions. In particular, with $Fe^{3+}$ content increase the samples exhibit a magnetic transition at 83 K, which is then split into two magnetic transitions at 77 and 86 K, and finally, the $Fe^{3+}$-rich sample $Co_2FeBO_5$ shows antiferromagnetic ($T_{N1}$=110 K) and ferrimagnetic-like ($T_{N2}$=70 K) transitions similar to the end compound $Fe_3BO_5$. An analysis of the Co-containing ludwigite systems indicates that the chemical substitution modifies the magnetically active subsystem only if it results in the appearance of magnetic ions at M4 site ($Co^{2+}$, $Mn^{3+}$, or $Fe^{3+}$).

Based on the low-temperature neutron diffraction, *Freitas et.al* [29] have found a rather small magnetic moment (~0.5 $\mu_B$) for Co ions occupying the M4 site, instead of the expected 4 $\mu_B$ for high-spin $Co^{3+}$ ($t_{2g}^4 e_g^2$, S=2), while the observed moments of Co at the M1 site (~3.6 $\mu_B$), at the M2 site (3.1 $\mu_B$), and at the M3 (3.8 $\mu_B$) are consistent with those expected for high-spin $Co^{2+}$ ions ($t_{2g}^5 e_g^2$, S=3/2). It was then assumed that $Co^{3+}$ ions in $Co_3BO_5$ are in the low-spin state ($t_{2g}^6 e_g^0$, S=0). It was found also that all magnetic moments are almost parallel to the *b*-axis, making it an easy magnetization direction. The spin ordering within 3-1-3 triad is antiferromagnetic (↑↓↑). These triads are ferromagnetically coupled with each other via Co ions at the M2 site resulting in the total uncompensated magnetic moment ~1.4 $\mu_B$ per Co ion. The recent measurements of crystal structure, magnetic susceptibility, electrical conductivity, and soft X-ray spectroscopy of $Co_3BO_5$ performed in the paramagnetic regime *T*=300-700 K, revealed the existence of two phase transitions at 475 and 495 K [30]. These transitions are well-defined from the differential scanning calorimetry (DSC) and electrical conductivity experiments and are accompanied by anomalies in the unit-cell parameters, although the type of crystal structure is retained and the space group remains unchanged (*Pbam*). A large paramagnetic moment 4.87 $\mu_B$ per Co ion was obtained from the fit of high-temperature part of magnetic susceptibility above room temperature and a deviation from Curie-Weiss behavior below this temperature was found. Although direct experimental evidence of spin transition is still missing, the authors of work [30] deduced that $Co_3BO_5$ undergoes a gradual $Co^{3+}$ spin-state crossover in the temperature interval $T_N \leq T < 300$ K and two-step changes in the Co oxidation state in each of four crystallographic sites at high temperatures. Indeed, the effective magnetic moment $\mu_{eff}$=4.16 $\mu_B$/Co obtained from the fitting of the susceptibility in the interval *T*=100-300 K [16, 18] can be compared with the value $\mu_{eff}$ = 4.24 $\mu_B$ expected for the case in which all Co ions, including the trivalent ones, are in high spin states (without orbital moment contribution). This observation indicates that a crossover or a transition from LS to HS could occur in $Co_3BO_5$. However, the study of thermodynamic properties of $Co_3BO_5$ below room temperature [16] did not reveal any anomaly except the magnetic one at $T_N$=42K. At the same time, conductivity and differential scanning calorimetry (DSC) measurements [30] performed on both single crystalline and powder samples at $T$>300 K showed anomalies, which are typically associated with the first-order phase transition. However, no such phase transition is observed in $Co_3BO_5$. Consequently, the magnetic properties and origin of the magnetic moments in the $Co_3BO_5$ remains unclear.

Thus, the subject of the present work is to determine the spin and oxidation states of cobalt ions in $Co_3BO_5$ and to study their evolution with temperature. We have performed single-crystalline X-ray diffraction, differential scanning calorimetry, and heat capacity measurements in a wide temperature range of 296-773 K. The Co $K$-edge X-ray magnetic circular dichroism (XMCD) and angle-dependent *dc* susceptibility experiments were carried out in the ferrimagnetic phase (~5 K) and above the magnetic phase transition (100-250 K), respectively. We calculated the electronic structure and magnetic moments of Co atoms at different metal sites for two paramagnetic phases: 296 K (low-temperature phase) 703 K (high-temperature phase) using the density-functional theory (DFT) with the account of strong Coulomb correlations through GGA+U scheme.

These combined experimental and theoretical studies indicate that at temperatures up to room temperature the magnetic lattice of $Co_3BO_5$ is formed mostly by divalent $Co^{2+}$ ions (S=3/2 with a small orbital contribution). The high-temperature investigation of crystal structure did not reveal any structural transformations (sp.gr. *Pbam*) whereas the lattice parameters and volume show a large thermal expansion and anomalies at $T_1$=500 and $T_2$=700 K. The unit cell volume increases by about 4.5 %. We found that the octahedral environment and the oxidation state of Co ions at the M4 site are drastically changed, while the oxidation state of other metal sites remains unchanged. The measurements of DSC indicate that the compound does not experience any first/second-order phase transitions besides those into the magnetically ordered state at $T_N$. This is confirmed by the thermodynamic properties measurements, which revealed the existence of two smeared anomalies in the high temperature range which are consistent with those observed in the X-ray diffraction experiments.

Our GGA+U calculations have shown, first of all, that the $Co^{2+}$ ions are in the high-spin state, while $Co^{3+}$ are in the low-spin state at low temperatures. Second, in the high temperature phase ($T$=703 K) the charge ordering disappears, all Co ions have the same $3d^7$ electronic configuration, and the system becomes metallic with all Co ions in the high-spin state. Moreover, our calculations point out that the spin-state transition and charge ordering does not necessarily occur at the same temperature.

## 2. EXPERIMENTAL TECHNIQUES

Needle-shaped single crystalline specimens of $Co_3BO_5$ have been synthesized using a flux method in the system $Bi_2Mo_3O_{12}$- $B_2O_3$- CoO - $Na_2CO_3$ - $Co_2O_3$ [18].
X-ray Absorption Near-edge Structure (XANES) and X-ray Magnetic Circular Dichroism (XMCD) measurements at the *K*-edges of Co and Fe have been performed at the ESRF ID12 beamline. The first harmonic of APPLE-II type helical undulator was used as a source of circularly polarized X-rays. To further monochromatize the undulator emission, a fixed-exit double crystal Si(111) monochromator was used at the required photon energies. The single crystals of $Co_3BO_5$ and $Co_2FeBO_5$ were mounted on a cold finger of an "amagnetic" He flow cryostat that was inserted in a cold bore of superconducting solenoid producing a maximum magnetic field of 17 Tesla. All XANES spectra were recorded using total fluorescence yield detection mode in "back-scattering" geometry using Si photodiodes mounted on the liquid nitrogen shield of the solenoid allowing very large solid angle of detection. Experiments were performed at 5 K under a magnetic field of ±17 T to ensure that the magnetic saturation was reached. Samples were oriented so that the direction of the magnetic field and incident x-ray wavevector were collinear with the crystallographic *b*-axis, which is an easy magnetization direction for both samples. The normalized XANES spectra were corrected for self-absorption effects taking into account the various background contributions (fluorescence of outer electronic subshells and other elements in the sample as well as coherent and incoherent scattering) and the solid angle of the detector. The XMCD signal was obtained as direct difference of XANES spectra measured with opposite helicities of the light at a fixed magnetic field. The resulting XMCD spectra are measured repeatedly with the opposite direction of magnetization to eliminate any artifacts.

For magnetic measurements a single crystal of 0.41 mg has been mounted in the rotator option of a SQUID-based magnetometer. The sample has been fixed with a pinch of vacuum grease with the rotator axis parallel and perpendicular to the needle axis (*c*-axis). We could select the different *a*-, *b*- and *c*- axis for the measurements as a function of temperature. DC Magnetic susceptibility at 1 kOe has been measured from 100 K to 250 K at different orientations. When rotating the sample at 200 K we observe a large anisotropy.

The simultaneous thermal analysis (STA) of $Co_3BO_5$ sample was performed using Jupiter STA 449C analyzer (NETZSCH, Germany) equipped with an Aëolos QMS 403C quadrupole mass spectrometer (NETZSCH, Germany) in Pt-Rh crucibles. A powdered sample of $Co_3BO_5$ single crystals with a weight of 160 mg was used. The measurements of mass change (TG) and heat flux (DSC) was performed in sequential heating and cooling mode at a rate of 10 °C/min within temperature range 373-773 K, supplying the different gas mixtures (20 vol.% $O_2$ in Ar and Ar with purity 99,9995 vol.%) with total flow rate of 50 $cm^3$(NTP)/min. Two successive heating-cooling cycles were carried out for each gas mixture. The first heating-cooling cycle was used for sample conditioning and the second cycle for data processing.

The specific heat capacity data from 373-773 K were obtained by the "ratio method" using a differential scanning calorimeter Netzsch STA Jupiter 449 C equipped with a special sample holder for $C_p$ measurements. Three different runs under the same conditions (dynamic argon-oxygen atmosphere 20 vol.% $O_2$, the heating rate 10 K/min) were carried out: (1) the baseline (empty Pt-Rh crucibles with perforated lids); (2) a standard sapphire disk (112 mg, with diameter of 6 mm and high of 1 mm) in the sample crucible; (3) the powder sample of $Co_3BO_5$ (60 mg) was manually sealed in the sample crucible. For enhancement of precise determination of heat capacity, it is advantageous when the "thermal masses" of the sapphire and the sample are approximately equivalent, i.e. $(m_{st} \times C_{p,st}) \approx (m_{sa} \times C_p)$.

The specific heat capacity of the sample was determined on the corrected Differential Scanning Calorimetry (DSC) curves according to equation:

$$C_p = \frac{m_{st}}{m_{sa}} \frac{(DSC_{sa} - DSC_{bl})}{(DSC_{st} - DSC_{bl})} C_{p,st} ;$$

where $C_{p,st}$ is the tabulated specific heat of the standard at temperature $T$; $m_{st}$, $m_{sa}$ are masses of the standard and the sample; $DSC_{sa}$, $DSC_{st}$ and $DSC_{bl}$ is the value of DSC signal at temperature $T$ from the sample, the standard and the baseline curve, respectively. The relative error of $C_p$ measurements did not exceed ±1%.

The low-temperature part of specific heat was measured by a relaxation technique in the temperature range 2-300 K on a single crystalline sample with a mass of 3.5 mg using a commercial instrument (Quantum Design PPMS).

The X-ray diffraction patterns were collected from a single crystal of $Co_3BO_5$ at 296 K, 403 K, 503 K, 603 K and 703 K using the SMART APEX II single crystal diffractometer (Bruker AXS, analytical equipment) equipped with a PHOTON 2 CCD-detector, graphite monochromator and Mo Kα radiation source. The measurements were carried out in the Common Access Facility Centre of SB RAS in Krasnoyarsk. The space group was defined by the analysis of extinction rules and intensity statistics obtained from all reflections. Multiscan absorption correction of reflection intensities was performed by SADABS program. Then, the intensities of equivalent reflections were averaged. The structure was solved by direct methods [31] using the SHELXS program. The structure refinement was carried out by least-square minimization in SHELXL program [32] using anisotropic thermal parameters of all atoms.

## 3. DETAILS OF THE CALCULATION

The electronic structure of $Co_3BO_5$ was calculated within the density-functional theory (DFT) using the Vienna Ab initio Simulation Package (VASP) [33]. The Perdew-Burke-Ernzerhof (PBE) version of the exchange-correlation potential [34] was utilized. The k-mesh with up to 225 points in the irreducible part of the first Brillouin zone was used. The total energy calculations were performed within generalized gradient approximation with account of on-site Coulomb repulsion (GGA+U) in rotationally invariant form [35]. The values of on-site Coulomb repulsion and Hund's exchange were taken to be $U = 7$ eV and $J_H = 0.9$ eV, respectively [36].

## 4. RESULTS AND DISCUSSION

### 4.1. X-ray Magnetic Circular Dichroism magnetometry

The element-selective XMCD spectroscopy allows to determine the magnetic behavior of the different elements in complex materials through the disentanglement of their contributions to the total magnetic response, or to compare the magnetic behavior of a chosen element in different materials. Here we focus on the isostructural compounds $Co_3BO_5$ and $Co_2FeBO_5$.

In $Co_2FeBO_5$ the $Fe^{3+}$ ions substitute $Co^{3+}$ ions at M4 sites as it was found from XRD and Mössbauer spectroscopy studies [18,37]. This material has a high magnetic uniaxial anisotropy with the *b*-axis as an easy magnetization direction inherent to $Co_3BO_5$ and undergoes Ferri- ($T_{N2}=70$ K) and AF ($T_{N1}=115$ K) magnetic transitions similar to $Fe_3BO_5$ revealing a great impact of the magnetic ion at the M4 site. The $Co_2FeBO_5$ displays a strong reduction of the remanent magnetization compared to $Co_3BO_5$. Assuming the low-spin state of $Co^{3+}$ one might expect a large uncompensated magnetic moment in 4-2-4 triad, and hence in all the unit cell (this is a crude approximation since there is never the situation where all $Co^{3+}$ ions are in LS), while the appearance of the ion with a nonzero magnetic moment at M4 site should result in the decrease of the moment under the condition of the antiparallel arrangement of this moment relative the host magnetic system of $Co^{2+}$. Indeed the remanent magnetization in $Co_3BO_5$ was found to equal $M_r \approx 72$ emu/g whereas in $Co_2FeBO_5$ the $M_r$ is reduced down to 20 emu/g [18]. Based on the neutron powder diffraction and the angle-dependent magnetization data the schematic representation of magnetic couplings in $Co_3BO_5$ and $Co_2FeBO_5$ can be presented as shown in Fig. 2.

The $Co^{3+}$ ion in low-spin state does not have unpaired spins ($t_{2g}^6 e_g^0, S = 0$) and, therefore, it should not contribute to the magnetic moment of $Co_3BO_5$. If this is the case, the $Co^{2+}$ ions, residing at the 3-1-3 triads and site 2 of the 4-2-4 triads, should resemble those in $Co^{2+}_2Fe^{3+}BO_5$ (this is a crude approximation since the $Fe^{3+}$ ion is magnetic and is certainly coupled to the $Co^{2+}$ sublattice in the 4-2-4 triads).

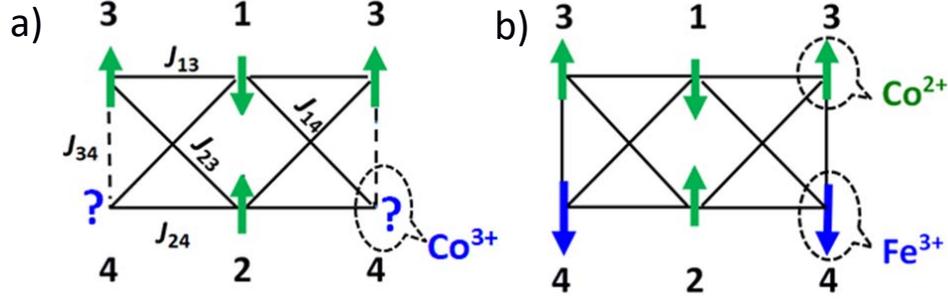

Fig. 2. Schematic representation of magnetic couplings in $Co_3BO_5$ (a) and $Co_2FeBO_5$ (b). The direction of $Co^{2+}$ and $Fe^{3+}$ magnetic moments, shown by the green and blue arrows, respectively, reflects the antiferromagnetic coupling between two subsystems. The magnetic moments of $Co^{3+}$ ions located at M4 site in $Co_3BO_5$ are not shown. The solid lines denote the superexchange interactions between the nearest magnetic ions $Co^{2+}$–O–$Co^{2+}$ and $Co^{2+}$–O–$Fe^{3+}$ ($J_{ij}$), while the dashed line is the magnetic coupling $Co^{2+}$–O–$Co^{3+}$. More detailed consideration on superexchange interactions in the ludwigites is presented in Ref. [38].

To directly address the problem of the spin state of $Co^{3+}$ ions, the XMCD experiment was performed at the Co $K$-edge under the same temperature and field conditions on both $Co_3BO_5$ and $Co_2FeBO_5$ single crystals. The fully ferrimagnetically ordered phases were considered. The XMCD spectra are shown in Fig. 3 along with corresponding normalized XANES spectra. The pre-edge peak which is sensitive to the orbital component of the magnetic moment was analyzed. This hard X-ray spectroscopy probes the transitions from a $1s$ core state to the empty $4p$ states and the pre-edge feature (7710-7715 eV) arises from transitions to more localized unoccupied $3d$ states resulting from the Co$4p$-$3d$ hybridization. While the shapes of Co($K$) XANES spectra are very similar for both samples a significant difference in their intensity is seen. This is related to the changes of local symmetry around Co atoms. Because the cobalt local symmetry and magnetization probed by XANES/XMCD provide information averaged over all non-equivalent crystallographic positions of Co, the shape and the intensity of the spectra measured at Co $K$-edge is expected to be dependent on the samples. The intensity of XANES pre-edge feature is reduced in $Co_2FeBO_5$ sample relative to that of the $Co_3BO_5$ indicating both the $4p$-$3d$ hybridization and the number of the unoccupied $3d$ states are different. A clear XMCD signal is visible at the pre-edge feature of Co. A nonzero XMCD signal reflects the magnetic polarization of the Co $4p$ band resulting from the intra-atomic exchange interaction with the $3d$ band.

The sign of the Co $K$-edge XMCD signal implies that in both $Co_3BO_5$ and $Co_2FeBO_5$ the magnetization of the Co sublattice is parallel to the applied magnetic field. The shape of the Co $K$-edge XMCD spectra is very similar for $Co_3BO_5$ and $Co_2FeBO_5$ while the signal amplitude at the pre-edge feature enhances as the $Co^{3+}$ ions are replaced by $Fe^{3+}$ ions. Then, we can assume that the Co *K*-edge XMCD pre-peak (7710-7715 eV) spectrum of $Co_3BO_5$ is composed of the $Co^{2+}$ component, $XMCD_{Co}^{2+}$, and a contribution from $Co^{3+}$, $XMCD_{Co}^{3+}$, while the XMCD signal of $Co_2FeBO_5$ depends on $XMCD_{Co}^{2+}$ component only:

$$XMCD(Co2Co) = \frac{1}{3} \cdot (2 \cdot XMCD_{Co^{2+}} + XMCD_{Co^{3+}})$$

$$XMCD(Co2Fe) = \frac{1}{2} \cdot (2 \cdot XMCD_{Co^{2+}})$$

The signal ratio is determined as follows

$$\frac{XMCD(Co2Co)}{XMCD(Co2Fe)} = \frac{2}{3} \cdot \left(1 + \frac{XMCD_{Co^{3+}}}{2 \cdot XMCD_{Co^{2+}}}\right)$$

If the $Co^{3+}$ ions in $Co_3BO_5$ are in the LS state, they should not play any significant role in the magnetic polarization of the Co(*sp*) band and, consequently, in the Co *K*-edge XMCD spectra. If this is the case, the signal ratio should be close to 0.66. Any magnetic contribution of the $Co^{3+}$ ions increases this ratio. The magnitude of the XMCD signal was determined by the integration of the XMCD spectrum in the range of the quadrupole transition, and the energy interval was 7705-7715 eV. The ratio of the signals obtained in this way gives the value of 0.70±0.01. This result shows that there is a rather small net $Co^{3+}$ magnetization in $Co_3BO_5$ and this provides an experimental support for the LS state of the $Co^{3+}$ ions. The small difference in signals can be explained also by the difference in the contributions of $Co^{2+}$ sublattice in the direction of applied magnetic field in the two studied compounds.

In the framework of the additivity model, the Co *K*-edge XANES spectrum of $Co_2CoBO_5$ is composed of a $Co^{2+}$ component, $XANES_{Co}^{2+}$, and a $Co^{3+}$ one, $XANES_{Co}^{3+}$. We can assume that the contribution of the $Co^{2+}$ to the Co *K*-edge XANES spectra is the same for $Co_2CoBO_5$ and $Co_2FeBO_5$ compounds, and by subtracting $XANES_{Co}^{2+}$ spectrum the contribution of the $Co^{3+}$ to the XANES spectrum of $Co_2CoBO_5$ can be extracted. The $XANES_{Co}^{3+}$ spectrum obtained in this way corresponds to the $Fe^{3+}$ atom substituted for $Co^{3+}$ (Fig. 4a). The following noticeable features can be distinguished in $XANES_{Co}^{3+}$ spectrum. The first one is pre-edge feature which is well resolved (the inset to Fig. 4a) and composes of two peaks at 7712.2 eV (A1) and 7714.8 (A2). Such a doublet structure was observed earlier in a wide range of XAS experimental results obtained on the perovskite-like cobaltites $LaCoO_3$ and $GdCoO_3$ [39, 40]. This pre-edge feature can be assigned to the 1*s*-3*d* transition, which is forbidden in the dipole approximation but it manifests itself due to the *p-d* orbital mixing and quadrupole contribution. The energy separation of about 2.6 eV is close to those 2.3 eV and 2.6 eV found in $LaCoO_3$ [39, 40] and $GdCoO_3$ [41], respectively. The next rather intensive feature arises at ~7723.3 eV (A3), followed by a sharp increase in the absorption corresponding to the main 1*s*-4*p* transition (A4) with energy position of 7732 eV. It is interesting to compare the spectrum of $XANES_{Co}^{3+}$ in $Co_3BO_5$ with the XANES at the Co *K*-edge for spinel $ZnCo_2O_4$ measured at the same temperature and field conditions (Fig. 4b) [42]. The latter compound is a known *p*-type conducting oxide material in which $Co^{3+}$ is incorporated at octahedral sites ($Co_{Oh}$). The Co(K-edge) XANES spectra of both materials are virtually identical showing the onset of the Co main absorption at the same energy (7732 eV). This observation clearly reveals the presence of trivalent cobalt in $Co_3BO_5$ ludwigite at low temperatures. Moreover, the shift of the main absorption to higher energies compared with that in $Co_2FeBO_5$ (7729 eV) reflects the difference in $Co^{3+}$ and $Co^{2+}$ valence states and comparable with so-called chemical shift usually observed in Co-oxides [42].

The Fe *K*-edge XANES/XMCD spectra recorded on $Co_2FeBO_5$ are shown in Fig. 5. XANES spectrum exhibits two main features: the pre-edge peak located at ~7115 eV, which is attributed to the $Fe^{3+}$ ions occupying the octahedrally coordinated sites, and the main edge at ~7130 eV, which dominates the XMCD signal. The shape of the XMCD spectrum is different from that observed in $Ho_{0.5}Nd_{0.5}Fe_3(BO_3)_4$ borate [43] and shows a four-peak structure: one negative and three positive ones. The peak intensity (7120 eV) reaches 0.25%. This value is about one order larger than that measured in rare-earth ferroborate and is comparable with the amplitude obtained for pure Fe foil [44], showing that Fe sublattice is nearly magnetically saturated in $Co_2FeBO_5$.

Another interesting feature from the XMCD study is that it provides a simple explanation to the reduction of the magnetic moment under the substitution of $Co^{3+}$ for $Fe^{3+}$. The element-selective magnetization curves recorded at Co and Fe K-edges present hysteresis loops with opposite signs (Fig. 6). This observation clearly indicates an antiferromagnetic coupling between Co and Fe subsystems and is in line with our previous XMCD measurements at the $L_{2,3}$ edges [45]. The Fe magnetization curve almost reaches saturation, while the Co magnetization curve has a slope highlighting the ferrimagnetic arrangement of cobalt magnetic moments in the $Co_2FeBO_5$. Besides the external magnetic field *H*, which determines the alignment for all magnetic moments and defines the positive direction, two exchange fields act on the Fe magnetic moments: the exchange field $H_{ex}^{Fe-Fe}$ created by Fe spins, which favors the parallel alignment of all Fe moments, and the $H_{ex}^{Fe-Co}$ created by the Co moments. Similar contributions are acting on the Co magnetic moments: *H*, $H_{ex}^{Co-Co}$, $H_{ex}^{Co-Fe}$. The fact that the Co magnetization has same orientation in both $Co_3BO_5$ and $Co_2FeBO_5$, i.e. positive with respect to the applied field *H*, indicates that the Co magnetic subsystem determines the trend in the magnetization of the ludwigites. In the ferrimagnetic phase, where exchange fields are larger than *H*, the $H_{ex}^{Fe-Co}$ tends to align the Fe moments antiparallel to the Co ones, and consequently, to *H*. This magnetic coupling seems to be the strongest if we assume that ordering temperature ($T_{N1}$) reflects the scale of exchange interactions. As a result, the bulk magnetization of the Fe-substituted ludwigite is significantly reduced due to partial compensation of Co and Fe magnetic moments.

The enhanced impact of Co contribution to the overall magnetization can be explained by the presence of a noticeable orbital moment $m_L$ of $Co^{2+}$ ion, which is only partially quenched. At low $T$ the orientation of $m_L$ is determined by both $H$ and the intra-atomic spin-orbit coupling $H_{LS}$, which, according to Hund's third rule, favors the parallel alignment of the orbital ($m_L$) and spin ($m_S$) moments of $Co^{2+}$ ion.

The experimental changes in the Co $K$-edge XMCD for two Co-containing ludwigites, supplemented by measurements at the Fe $K$-edge, were found to correlate with the overall magnetism of the system $Co_{3-x}Fe_xBO_5$ ($x$=0.0 and 1.0) and qualitatively agree quite well with the magnetic order proposed in Fig. 2. Our results underline the key role of the M4 ion because it is the filling of this site by different magnetic ions, which leads to the dramatic modification of the magnetic ground state. The $XANES_{Co}^{3+}$ spectrum reveals features and onset of Co main absorption similar to those observed in spinel $ZnCo_2O_4$ providing the evidence of trivalent cobalt in $Co_3BO_5$. We find that it is better to view these ludwigites in terms of $Co^{2+}$ and $M^{3+}$=Co, Fe sublattices. In summary, the XMCD measurements have shown that the magnetism of $Co^{2+}$ sublattice has a similarity in two isostructural compounds $Co_3BO_5$ and $Co_2FeBO_5$, and the contribution from $Co^{3+}$ ion is strongly suppressed in $Co_3BO_5$.

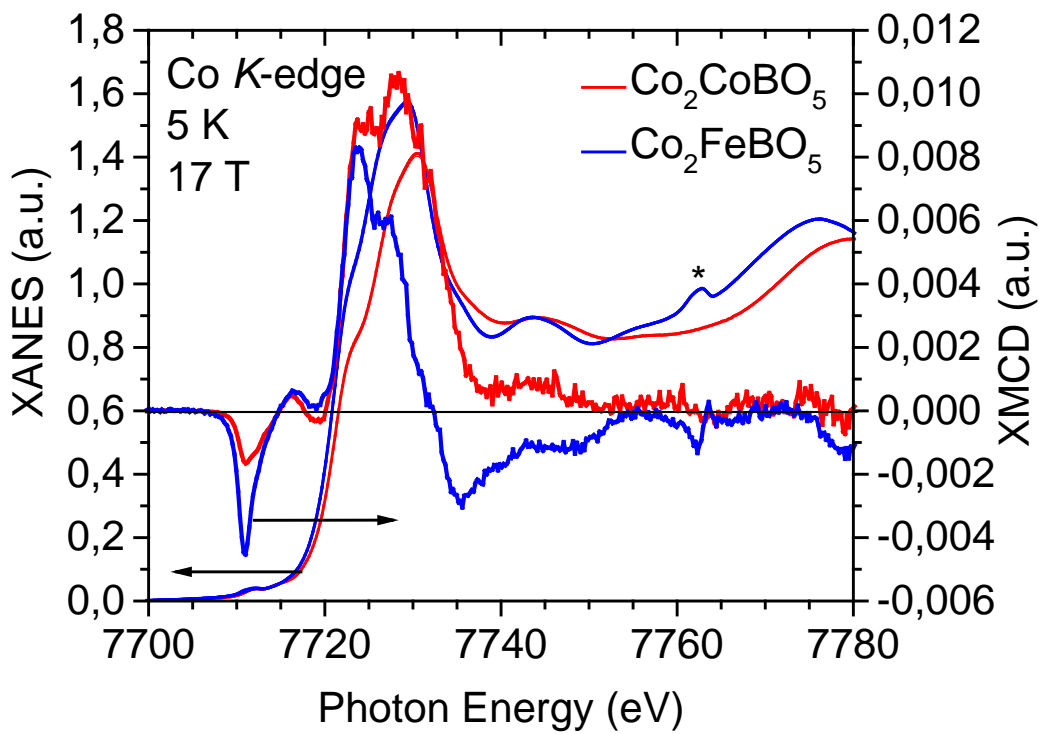

Fig. 3. Normalized Co $K$-edge XANES/XMCD spectra recorded at $T$=5 K and applied field 17 T for $Co_3BO_5$ and $Co_2FeBO_5$ single crystals. The asterisk denotes the diffraction peak from the crystal.

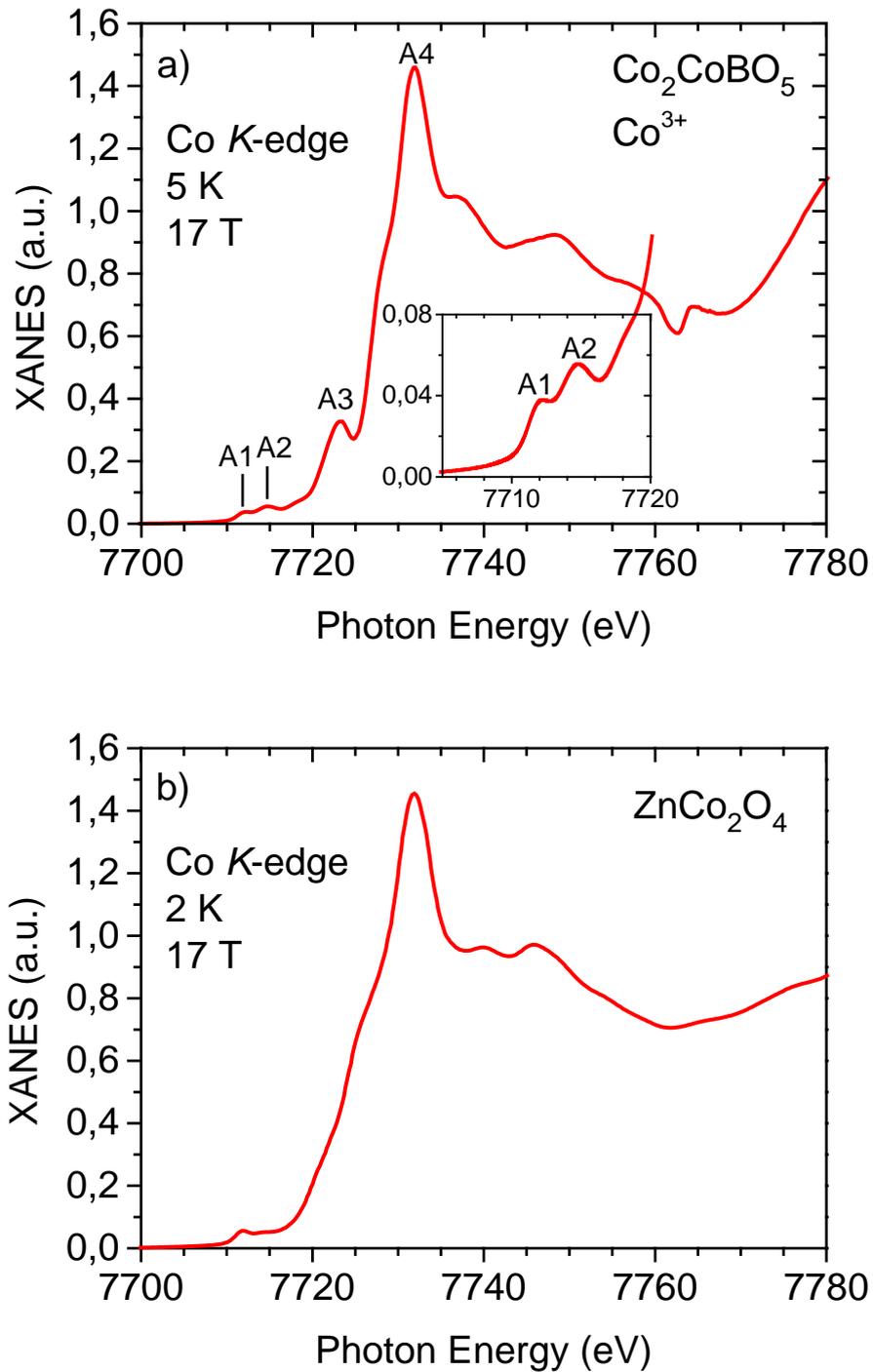

Fig. 4. a) XANES spectrum corresponding to the Co$^{3+}$ ion in Co$_3$BO$_5$ obtained by the subtracting of 2*XANES spectrum of Co$_2$FeBO$_5$ from 3*XANES spectrum of Co$_3$BO$_5$. b) XANES spectrum recorded at the Co $K$-edge for the spinel ZnCo$_2$O$_4$, data are taken from the Ref. [42].

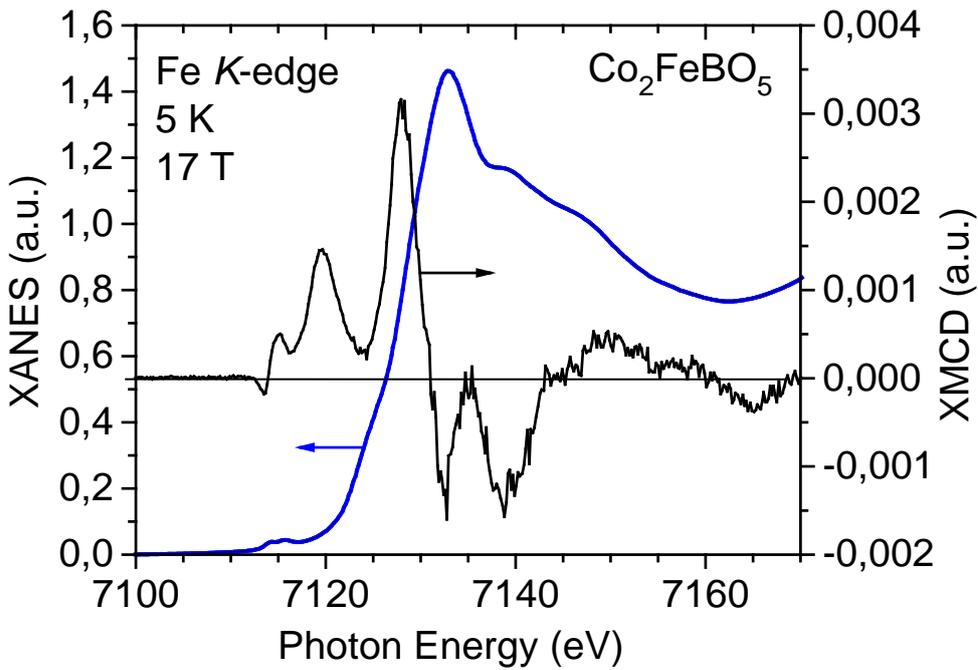

Fig. 5. Normalized Fe *K*-edge XANES/XMCD spectra for $Co_2FeBO_5$ single crystals, *T*=5 K and applied field is 17 T

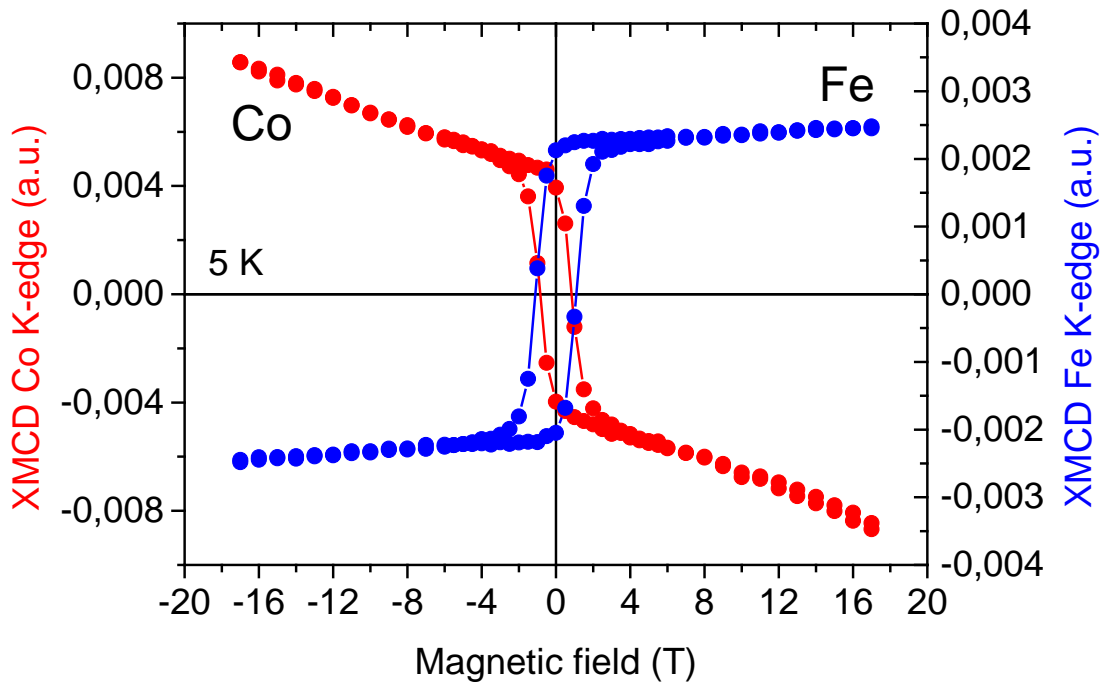

Fig. 6. The element-selective magnetization curves of $Co_2FeBO_5$ single crystal recorded at Co *K*- and Fe *K*-edges for applied field along the EMD (*b*-axis) and *T*=5 K.

## 4.2. Paramagnetic contribution of $Co_3BO_5$ as a function of crystal orientation

In order to study the $Co^{3+}$ contribution in magnetism of $Co_3BO_5$ in the paramagnetic state we have performed orientation-dependent susceptibility measurements on $Co_3BO_5$ needle shaped single crystal at temperature range 100-250 K.

Results of the magnetic characterization measurements are presented in Fig. 7. The paramagnetic susceptibility of $Co_3BO_5$ is highly anisotropic, where the *b*-axis is an easy magnetization direction (EMD) and the *c*-axis is a hard magnetization one (HMD). The measurement in green has been done without the rotator with the needle perpendicular to field (*H* in the *ab* plane). The χ·T is relatively flat for this measurement. The measurements with the rotator have some drift which can be due to the inherent error in the background measurement which presents some orientation dependence (7-10 % deviations). This is why these curves were fitted taking into account a temperature independent contribution (inset to Fig. 7).

The temperature dependence of the magnetic susceptibility in the paramagnetic state obeys the Curie–Weiss law,

$$\chi = \chi_0 + \frac{C}{T-\theta} \qquad (1)$$

Fitting the data yields a temperature independent term $\chi_0$, the Curie–Weiss constant *C*, and the Curie–Weiss temperature $\theta$ collected in Table 1.

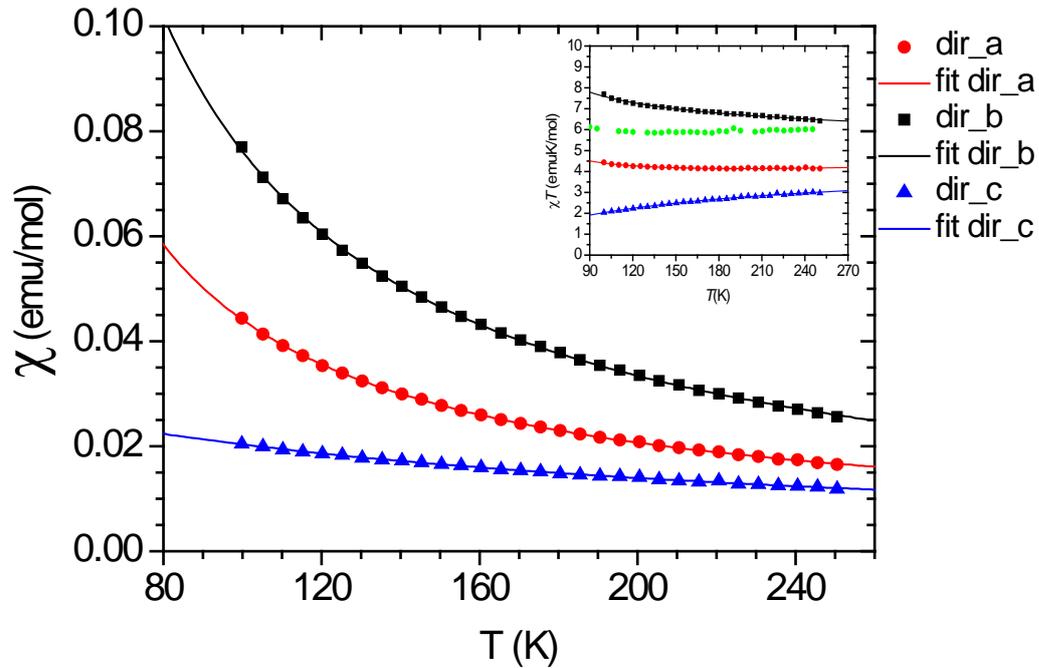

Fig. 7. Temperature dependence of magnetic susceptibility for $Co_3BO_5$ single crystal in applied field oriented parallel the *a*, *b*, and *c*-axis. Symbols denote the experimental data, the straight lines are fitting using Eq.1. Inset: χT vs T which fitted to Eq.1

Table 1. Magnetic parameters of $Co_3BO_5$ extracted in the paramagnetic phase.

|  | $\chi_0$ (emu/mol) | C (emu·K/mol) | θ (K) | $\mu_{eff}$ ($\mu_B$/f.u.) |  |
|---|---|---|---|---|---|
| *a*-axis | 2.4±0.2 ·10⁻³ | 3.3±0.1 | 22±1 | 5.1±1 | IMD |
| *b*-axis | -0.6±0.2 ·10⁻³ | 6.1±0.1 | 20±1 | 7.0±0.1 | EMD |
| *c*-axis | -0.2±0.1 ·10⁻³ | 4.6±0.5 | -122±16 | 6.0±0.1 | HMD |

The positive sign of the Curie-Weiss temperature indicates the predominance of ferromagnetic exchange interactions between Co ions in the *a*- and *b*- directions, while a strong negative interaction persists inside magnetic chains propagating along the *c*-axis. The Curie–Weiss constant *C* reflects strong anisotropy in the PM state. The average value of the constant is $C_{av} = \frac{C_a + C_b + C_c}{3} = 4.7 \pm 0.1$ (emu·K/mol) yields an effective magnetic moment $\mu_{eff} = 6.1 \pm 0.1$ $\mu_B$ per formula unit, which corresponds to a moment of 3.53±0.06 $\mu_B$ per Co ion. This value is much lower than the previously reported ones [16, 18].

It is obvious that the magnetic moment of $Co_3BO_5$ depends on the number of magnetically active ions $n_{Co^{2+}}$, $n_{Co^{3+}}$, on the *g*-factors $g_{Co^{2+}}$ and $g_{Co^{3+}}$ and on the spin values $S_{Co^{2+}}$ and $S_{Co^{3+}}$ of $Co^{2+}$ and $Co^{3+}$ ions, respectively:

$$\mu_{eff}^S = \sqrt{\frac{n_{Co^{2+}} \cdot g_{Co^{2+}}^2 \cdot S_{Co^{2+}} \cdot (S_{Co^{2+}} + 1) + n_{Co^{3+}} \cdot g_{Co^{3+}}^2 \cdot S_{Co^{3+}} \cdot (S_{Co^{3+}} + 1)}{n_{Co^{2+}} + n_{Co^{3+}}}} \quad (2)$$

Assuming that all magnetic ions are in high-spin state, the spin-only magnetic moment comprises of magnetic moments of $Co^{2+}$ ($n_{Co^{2+}}=2$) and $Co^{3+}$ ($n_{Co^{3+}}=1$) ions with spin values of $S_{Co^{2+}}^{HS}=3/2$ and $S_{Co^{3+}}^{HS}=2$, and $g_{Co^{3+}} = g_{Co^{2+}} = 2$. This gives an effective spin moment of $\mu_{eff}^S=4.24$ $\mu_B$/Co (right panel of Fig. 8). If the $Co^{3+}$ ions are in the LS state ($S_{Co^{3+}}^{LS}=0$) and the $Co^{2+}$ ions are still in HS state the magnetic moment is reduced to $\mu_{eff}^S=3.16$ $\mu_B$/Co, which is rather close to the experimental value. Moreover, if only two $Co^{2+}$ ions are contributing in magnetism of $Co_3BO_5$, i.e. $n_{Co^{2+}}=2$ and $n_{Co^{3+}}=0$, the experimentally observed effective spin moment per $Co^{2+}$ ion would be 4.31 $\mu_B$, very close to obtained values in the literature for $Co^{2+}$ in octahedral sites (4.7 – 5.2 $\mu_B$) where an orbital contribution is rather important [46]. Usually, for divalent cobalt ions the observed values of the $g_{Co^{2+}}$ are significantly larger than 2 due to the orbital contribution. The experimental values of the magnetic moments $\mu_{eff}$ for pyroborate $Co_2B_2O_5$ ($n_{Co^{2+}}=2$) [47], ludwigites $Co_{2.5}Ti_{0.5}BO_5$ [27] and $Co_{2.5}Sn_{0.5}BO_5$ ($n_{Co^{2+}}=2.5$) [28], kotoite $Co_3B_2O_6$ ($n_{Co^{2+}}=3$) [48], and $Co_4B_6O_{13}$ [49], which contain only $Co^{2+}$ ions as a source of magnetism as a function of the $n_{Co}$ per formula unit are shown in Fig. 8. All of them show the values of $\mu_{eff}$ very close to each other and to 4.9 $\mu_B/Co^{2+}$, which corresponds to $g_{Co^{2+}} \approx 2.5$. The assumption of spin-orbital contribution for $Co^{2+}$ ions with *g* factor obtained above will change the effective magnetic moment of $Co_3BO_5$ in two ways: i) the moment has a slow increment with the increase in the $n_{Co}$ if $Co^{3+}$ ions are assumed to be in HS state (orange line and values above it); ii) the moment rapidly decreases if $Co^{3+}$ ions are in LS state (blue line and values below). As can be seen the magnetic moment of $Co_3BO_5$ obtained at present work and in previous studies for more extended interval T≤300 K agrees well with the assumption of LS state of $Co^{3+}$. Moreover, if our assumption is valid, strong ferromagnetism of $Co_3BO_5$ with magnetic moment $M_r=3.4(1)$ $\mu_B$/f.u. should be attributed to the almost collinear ferrimagnetic ordering of two $Co^{2+}$ moments with g = 2.5 that is 0.91 of the expected $M_{sat} = ngS\mu_B$=3.75 $\mu_B$ per formula unit. So, one can conclude that at low temperatures (T<300 K) the value of effective magnetic moment (6.1 $\mu_B$/f.u.) is compatible with two $Co^{2+}$ ions in the HS state (and some orbital contribution) and the $Co^{3+}$ ions in a LS state. This fact additionally supports the results of neutron powder diffraction that only the $Co^{2+}$ ions are magnetic and the $Co^{3+}$ ions are in the LS state.

Note, that for $Co_{2.4}Ga_{0.6}BO_5$ and $Co_{2.88}Cu_{0.12}BO_5$ ludwigites whose magnetic moments contain the $Co^{2+}$ and a certain amount of $Co^{3+}$, the $\mu_{eff}$ falls into the range of LS values. Qualitatively, both compounds show a behavior similar to $Co_3BO_5$ reflecting a ferrimagnetic ordering of cobalt magnetic moments near 40 K [20,21], thereby indirectly pointing out the similar nature of Co magnetism in these compounds.

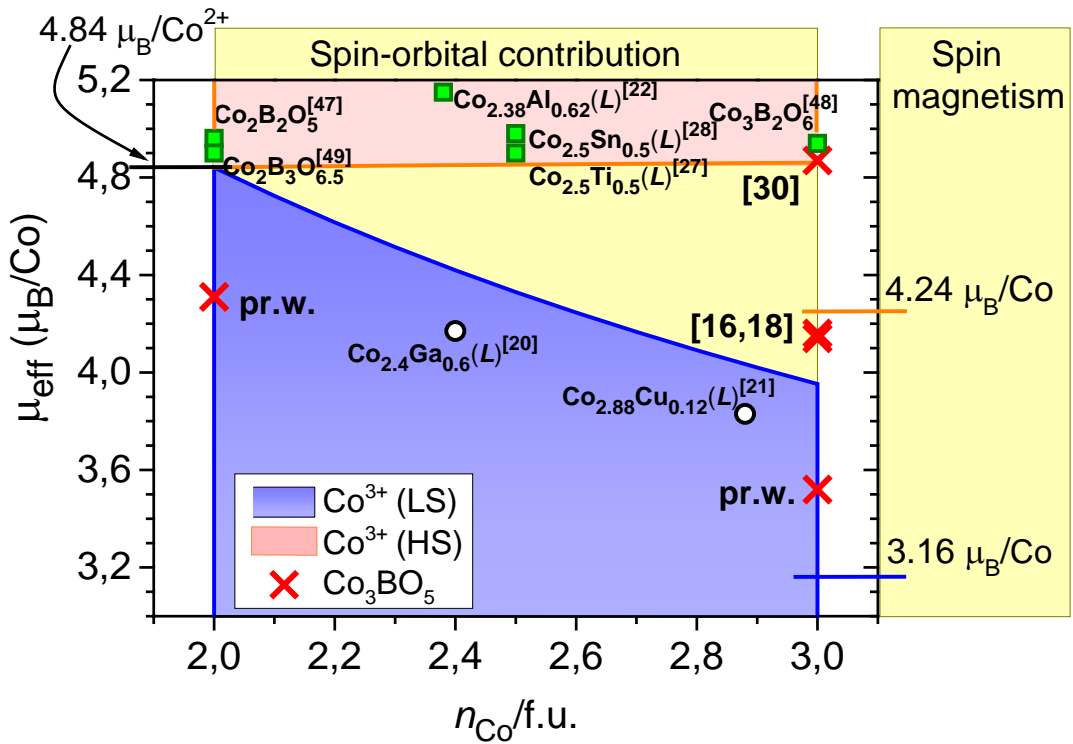

Fig. 8 Magnetic moment of Co-based borates/oxyborates *vs* number of cobalt ions per formula unit (see text). For a better comparability, the magnetic moment data are given per one cobalt ion. The symbol (*L*) denotes the *ludwigite*. Right panel shows the values of magnetic moments per Co ion in $Co_3BO_5$ expected in the spin-only magnetism approximation. The central panel contains the values of magnetic moments assuming some orbital contribution for HS $Co^{2+}$ ions: the orange line and values above are expected values for the HS $Co^{3+}$ ions, while the blue line and values below are for the LS $Co^{3+}$ ions in the Co-containing ludwigites. The red crosses denote the experimentally observed values of $\mu_{eff}$ for $Co_3BO_5$.

**4.3. Differential scanning calorimetry**

The mass change (TG) and heat flux (DSC) measurements of $Co_3BO_5$ are shown in Fig. 9. The DSC-TG curves indicate that there are no changes in weight or thermal effects in a wide temperature range 373-773 K. These results show that $Co_3BO_5$ is thermally stable in the temperature range of interest. Our observation is in strong contradiction with the data of Ref.

[30], where two endothermic (exothermic) peaks at 472 and 493 K were found under heating (cooling) cycles and assigned to phase transitions.

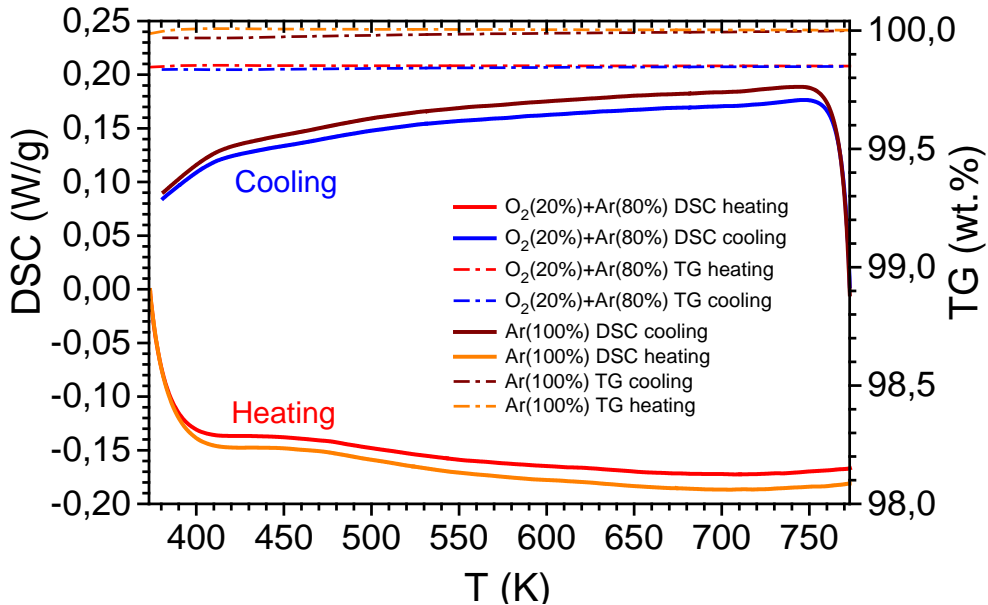

Fig. 9. DSC-TG curves (heating-cooling cycles) showing the thermal stability of the $Co_3BO_5$ powder sample in different gas mixtures: 20 vol.% $O_2$ in Ar and high pure Ar.

**4.4. Specific heat capacity**

The temperature dependence of the heat capacity taken at $H=0$ T is shown in Fig. 10a. At low temperatures, a quite pronounced λ-type anomaly at $T_N$=42 K implies a second-order phase transition which is in good agreement with the previously published heat capacity data [16]. In magnetic field the λ-type anomaly is progressively smeared out and shifts to higher temperatures (not shown). The above results combined with differential scanning calorimetry measurements confirmed that the title material does not exhibit a phase transitions at high temperatures in contrast to previous report [30].

For $Co_3BO_5$, the thermodynamic limit of the lattice contribution to the entropy $3Rz$=224.37 J/mol K, with $R$=8.314 J/mol K being the gas constant and $z$=9 the number of atoms per formula unit, is apparently reached at 440 K. Above this temperature the specific heat $C_p$ shows a significant contribution to the heat capacity, obviously of electronic or/and magnetic origin. To estimate the anomalous contribution of the specific heat $\Delta C_p$, we have fit the main phononic contribution to $C_p(T)$ using Debye-Einstein approximation at temperatures well outside the region of the $T_N$ anomaly. We obtain a reasonable value of Debye temperature $\Theta_D$=493±20 K. Note, the obtained value is larger than the 140 K found in the work [16], but it is in good agreement with the values extracted for other mixed-valence oxyborates $Mn_2BO_4$ ($\Theta_D$=512 K) [50] and $V_2BO_4$ ($\Theta_D$=360 K) [51] reflecting the rigid frame of chemical bonds in borate structures.

The anomalous contribution $\Delta C_p$ is shown in Fig. 10b. A significant anomalous contribution besides that at $T_N$ is seen in $\Delta C_p$ vs $T$ at elevated temperatures. This quite pronounced anomaly has not a jumplike shape, but a smeared one with a maximum at ~700 K and a clear shoulder at about 500 K. The best fitting of $\Delta C_p$ is obtained by the sum of two Gaussians centered at 526 K and 695 K.

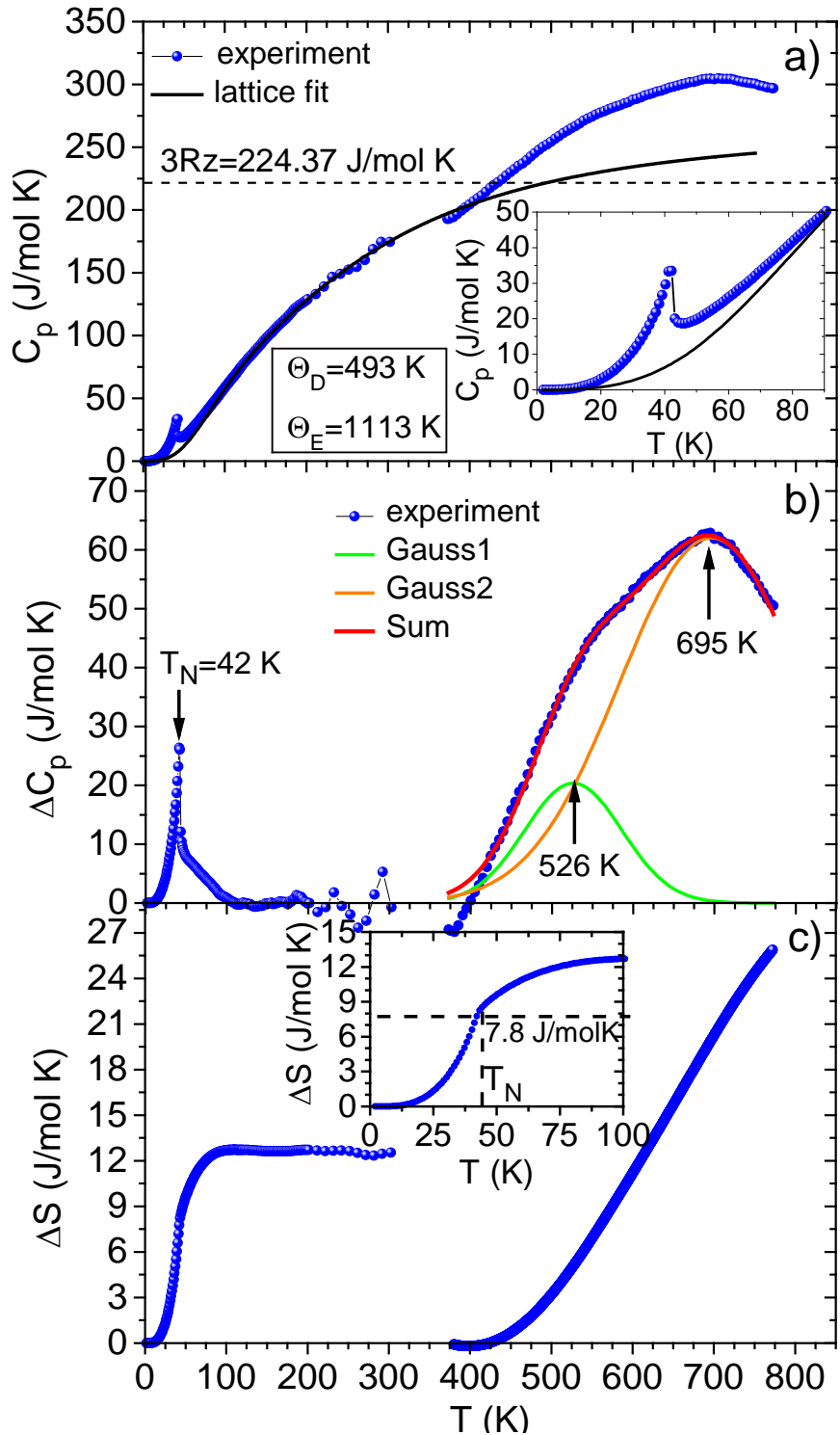

Fig. 10. a) Temperature dependence of the specific heat in $Co_3BO_5$ measured from 2 to 300 K and from 373 to 773 K. Dotted line indicates the Dulong-Petit value. The black line refers to a phonon contribution which has been fit to the data well outside the anomaly. Inset: enlarged low temperature part highlights the onset of ferrimagnetic spin ordering at $T_N$. b) The difference of the experimental data and the phonon contribution. Besides the $T_N$ the arrows show the anomalies in specific heat obtained from the sum of two Gaussians (green and orange). c) Entropy as a function of temperature. The inset is the range of magnetic phase transition.

The low temperature λ–peak entropy and enthalpy content has been calculated by extrapolating its HC below $T_N$ to a $T^3$ dependence, as corresponds to the spin-wave contribution of a ferrimagnetic lattice, and above $T_N$ to the high-temperature $T^{-2}$ dependence, yielding to the values:

   a) $0<T<T_N$, $\Delta S/R=1.00\pm0.2$, $\Delta H/R=33\pm 3$ K
   b) $T_N<T<\infty$, $\Delta S/R=0.7\pm0.2$, $\Delta H/R=59\pm3$ K

and total anomalous values $\Delta S/R=1.6\pm0.2$. We observe in Fig. 10c that the anomalous entropy has reached a plateau value already at 100 K. The total enthalpy is $\Delta H/R=92\pm 3$ K, expressed in gas constant units. The critical entropy: i.e. below $T_N$, per Co ion is $S_c(Co)/R=0.50\pm0.02$. The entropy above $T_N$ per $Co^{2+}$, $(S\infty-S_c)(Co)/R=0.68/2=0.34$, $(S\infty-S_c)/S_c=0.68$. The enthalpy is referenced to $U(T=\infty)=0$, then $(U\infty-U_o)(Co)/T_N=1.10$, $(U\infty-U_c)(Co)/T_N=0.70$. These values are comparable to those values found in Heisenberg antiferromagnets, for example to the SmGa Garnet [52].

The total anomaly entropy content is provided by the long range ordering of the magnetic $Co^{2+}$ ions, since the $Co^{3+}$ ions bear no magnetic moment. At this temperature the compound is an insulator, thus the $Co^{2+}$ electronic states are governed by crystal field interaction, and Co-Co exchange interactions. The former splits the $3d^7$ levels into Kramers doublets, and consequently, its ground state is a Kramers doublet [53, 54], with the excited level several hundred K higher energy. Below $T_N$, the latter, exchange interaction, further splits the ground doublet upon magnetic ordering. Therefore, the total entropy content expected, per ion, is $\Delta S_{ion}/R=\ln 2= 0.692$. Consequently, the experimental entropy content $\Delta S/R=1.6\pm0.2=(2.3\pm0.2)\ln2$, indicates that there are about 2.3 $Co^{2+}$ per formula unit in the $Co_3BO_5$ compound. Thus, within the experimental error, this result reinforces the conclusion that the M4 site is filled mostly by non-magnetic $Co^{3+}$, and the other sites are filled by the $Co^{2+}$ ($3d^7$) ions.

**4.4. High-temperature X-ray diffraction**

Several experimental studies of the $Co_3BO_5$ crystal structure have been performed [6,16,30,38,55]. Most of them consist of a structural characterization at room temperature and only works [16] and [30] are devoted to the study of temperature effects. In the latter the structural properties were studied using high-temperature X-ray powder diffraction. Here, we have performed the single-crystal X-ray diffraction measurements as a function of temperature that allows determining the temperature changes in the bond-lengths and the local coordination of the Co ions.

The structure determination was done at 296 K, 403 K, 503 K, 603 K, and 703 K. No conformation change was observed. The main information about crystal data, data collection and refinement are reported in Table 2. Coordinates of atoms are presented in Table 1S, while bond lengths are summarized in Table 2S of Supplemental Material [55]. The room-temperature lattice parameters are similar to those found in references [6,16,30,38,56]. For better comparability, the structural data obtained in the present work are superimposed on the data from Ref. 31 (Fig. 1S [55]). As can be seen, at room temperature the $a$-parameter shows a slight decrement compared with the reference data, which, however, decreases with temperature. Other lattice parameters are in good agreement with those referred. The observed difference in the $a$-parameter can be attributed to a small amount of metal vacancies, probably due to the peculiarities of the synthesis.

Table 2. Crystallographic data and main parameters of processing and refinement $Co_3BO_5$.

| | | | | | |
|---|---|---|---|---|---|
| Crystal data | | | | | |
| $M_r$ | | | 267.60 | | |
| Space group, Z | | | $Pbam$, 4 | | |
| Size, mm | | | 0.3×0.2×0.1 | | |
| T, K | 296 | 403 | 503 | 603 | 703 |
| $a$, (Å) | 9.2742 (5) | 9.2694 (4) | 9.2520 (7) | 9.2487 (17) | 9.2676 (11) |
| $b$, (Å) | 11.9590 (7) | 11.9902 (6) | 12.0745 (9) | 12.169 (2) | 12.2473 (15) |
| $c$, (Å) | 2.9787 (2) | 2.9906 (1) | 3.0167 (2) | 3.0314 (6) | 3.0467 (4) |
| $V$, (Å$^3$) | 330.37 (3) | 332.38 (2) | 337.01 (4) | 341.18 (11) | 345.81 (7) |
| $D_x$, Mg/m$^3$ | 5.380 | 5.348 | 5.274 | 5.210 | 5.140 |
| $\mu$, mm$^{-1}$ | 14.770 | 14.681 | 14.479 | 14.302 | 14.111 |
| Data collection | | | | | |
| Wavelength | | | MoK$_\alpha$, $\lambda$=0.7106Å | | |
| Measured reflections | 6883 | 6526 | 6754 | 6734 | 7355 |
| Independent reflections | 1058 | 914 | 953 | 938 | 1073 |
| Reflections with I>2σ(I) | 848 | 735 | 728 | 715 | 713 |
| Absorption correction | | | Multiscan | | |
| $R_{int}$ | 0.0697 | 0.0736 | 0.0776 | 0.0832 | 0.0968 |
| $2\theta_{max}$ (°) | 78.31 | 72.36 | 73.04 | 72.20 | 75.88 |
| h | -15 → 15 | -15 → 15 | -15 → 15 | -15 → 15 | -15 → 15 |
| k | -20 → 20 | -19 → 19 | -20 → 20 | -20 → 20 | -21 → 20 |
| l | -5 → 5 | -4 → 4 | -5 → 5 | -5 → 5 | -5 → 5 |
| Refinement | | | | | |
| $R[F^2>2\sigma(F^2)]$ | 0.0337 | 0.0309 | 0.0305 | 0.0333 | 0.0389 |
| $wR(F^2)$ | 0.0931 | 0.0657 | 0.0716 | 0.0738 | 0.0837 |
| S | 0.823 | 1.078 | 1.084 | 1.107 | 1.107 |
| Weight | w=1/[$\sigma^2(F_o^2)$+ (0.0808P)$^2$+ 2.653P] | w=1/[$\sigma^2(F_o^2)$+ (0.0244P)$^2$+ 0.644P] | w=1/[$\sigma^2(F_o^2)$+ (0.03214P)$^2$+ 0.380P] | w=1/[$\sigma^2(F_o^2)$+ (0.0280P)$^2$+ 0.513P] | =1/[$\sigma^2(F_o^2)$+ (0.0333P)$^2$+ 0.796P] |

| | | | where $P=\max(F_o^2+2F_c^2)/3$ | | |
|---|---|---|---|---|---|
| Extinction | 0.039 (3) | 0.040 (2) | 0.044 (2) | 0.045 (3) | 0.054 (3) |
| $(\Delta/\sigma)_{max}$ | <0.001 | <0.001 | <0.001 | <0.001 | <0.001 |
| $\Delta\rho_{max}$, e/Å$^3$ | 2.343 | 1.687 | 1.468 | 2.343 | 2.420 |
| $\Delta\rho_{min}$, e/Å$^3$ | -0.817 | -1.243 | -1.258 | -0.817 | -1.411 |

Note, that with increasing temperature, all parameters behave similarly to those obtained in Ref. 30, demonstrating that the slope changes at about 500 K. Additionally, the *a*-lattice parameter shows negative thermal expansion (Fig. 11). The first anomaly at $T_1$=500 K is clearly seen in the temperature dependences of the *a*, *b*- and *c*-lattice parameters and unit cell volume. With heating up the increase in the volume, $\Delta V/V$, reaches a value of 4.5%. The thermal expansion coefficients show an initial linear regime between 300 and 400 K. Above 400 K the three coefficients (*b*, *c*, and volume) rapidly increase reaching a maximum at about $T_1$=500 K and then grow again at $T_2$= 700 K.

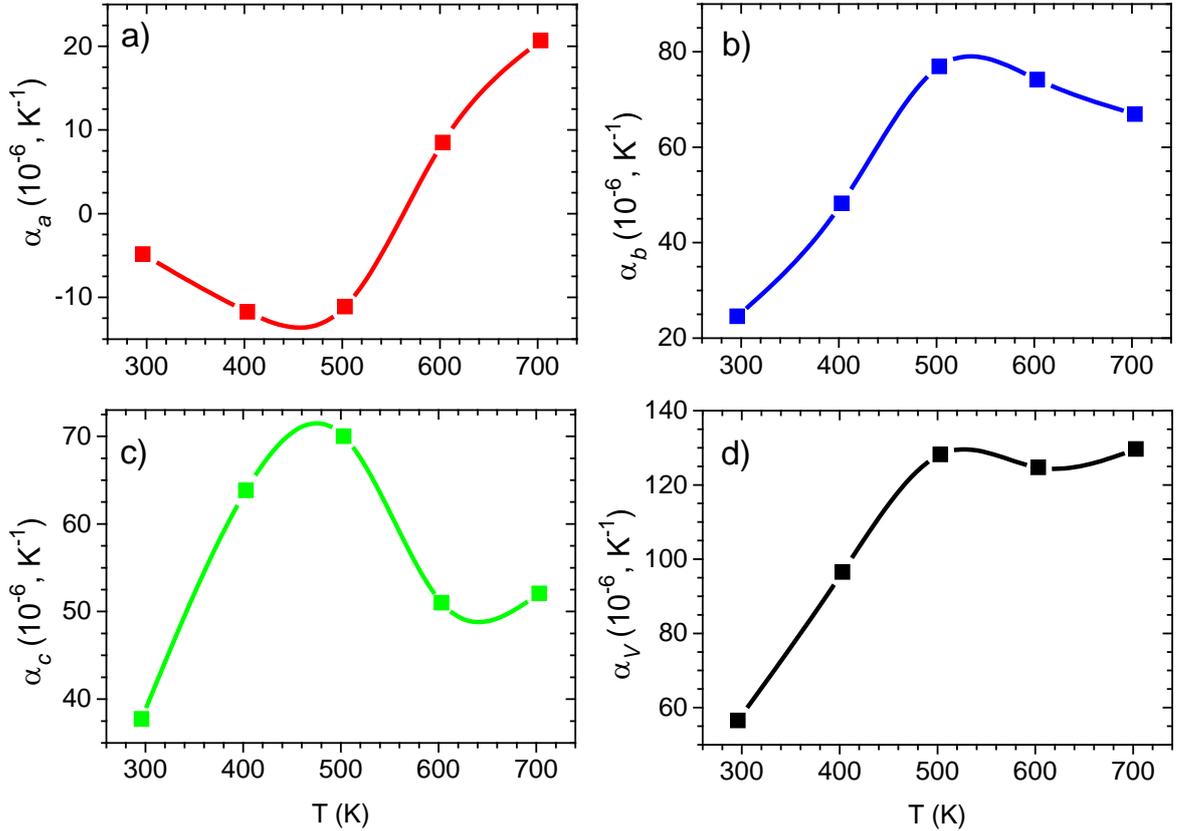

Fig. 11. Temperature dependence of the expansion coefficients: a) *a*-parameter, b) *b*-parameter, c) *c*-parameter, d) unit cell volume.

Similar to the lattice parameters, Co-O bond-lengths in $Co_3BO_5$ show an increase with temperature (Fig. 12a). The temperature has a dramatic effect on the $Co2O_6$ and especially on the $Co4O_6$ octahedra resulting in the rapid growth of the bond-lengths. The mean bond-lengths of Co1 and Co3 sites practically do not change with an increase in the temperature. For the analysis of the local distortions, we used the $V_{zz}$ parameter, which is the main component of the $G_{\alpha\beta}$ tensor of the electrical field gradient (EFG) produced by the different octahedral oxygen coordinations around the individual cobalt sites [26]. In Fig. 12b the main components $V_{zz}$ for different cobalt sites are plotted as a function of temperature. To begin with, we consider the local distortions at room temperature. The $Co^{2+}$ ions at Co1 and Co3 sites have a [2+4] coordination, forming axially compressed octahedra and, thus, showing the largest values of $V_{zz}$. Although the $Co2O_6$ shows axial compression, it is much less than that for $Co1O_6$ and $Co3O_6$. $Co4O_6$ octahedron has two short (equatorial) Co4-O1, two long (equatorial) Co4-O4 and two intermediate (axial) Co4-(O2)O3 bonds, thereby showing a slight axial elongation. The coordination bond-lengths seem to be more homogeneous around Co4 making it the most regular octahedron with smallest $V_{zz}(4)$.

The difference in the local deformations arises from different boron coordination. As it is seen from Fig. 13 for Co1 and Co3 sites four oxygen atoms in the equatorial plane (O3 for Co1 and O2, O5 for Co3) participate in short B-O bonds resulting in considerable elongation of appropriate bonds. However, there are only two oxygen atoms coordinated with the boron atoms in case of the $Co2O_6$ and $Co4O_6$ octahedra (O5 for Co2 and O2, O3 for Co4) also causing elongation of the Co4-O bond in direction of the $(BO_3)^{3-}$ groups. Note, the compression of $Co2O_6$ octahedron along Co2-O5 bonds is not in line with the expected distortion. The source of this discrepancy is probably the O4 atom in the equatorial plane, which is the bridge between Co2 and Co4 atoms and participates in the simultaneous compression of the Co2-O4 and the stretching of Co4-O4 bonds. This oxygen atom has relative freedom for the displacement, which facilitates the entering of the substitution atoms (Cu, Fe, Mn, Ga, Al) into the 4-2-4 triad [20,21,22,23,26].

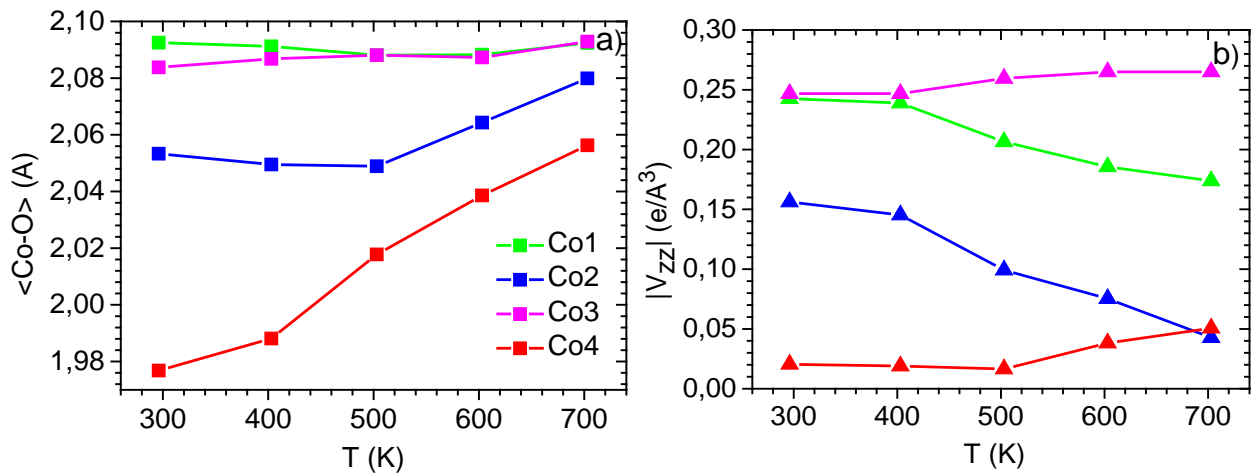

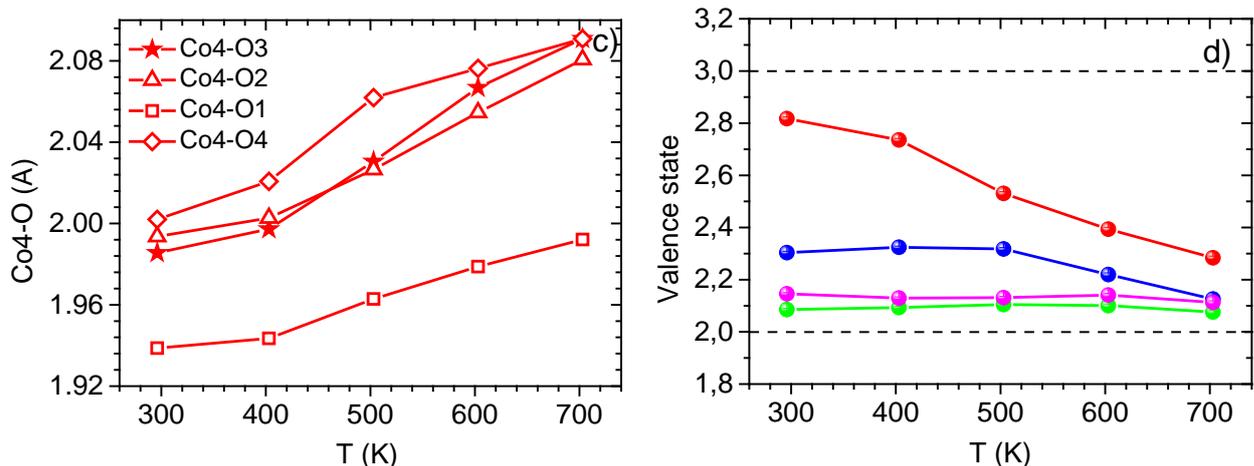

Fig. 12. Temperature dependences of mean bond-lengths (a), main components of EFG (b), octahedral bond-lengths in Co4O$_6$ octahedron (c), and bond-valence sum (average among Co$^{2+}$ and Co$^{3+}$) for cobalt ions in Co$_3$BO$_5$ (d). Dashed lines show integer valence states 2+ and 3+.

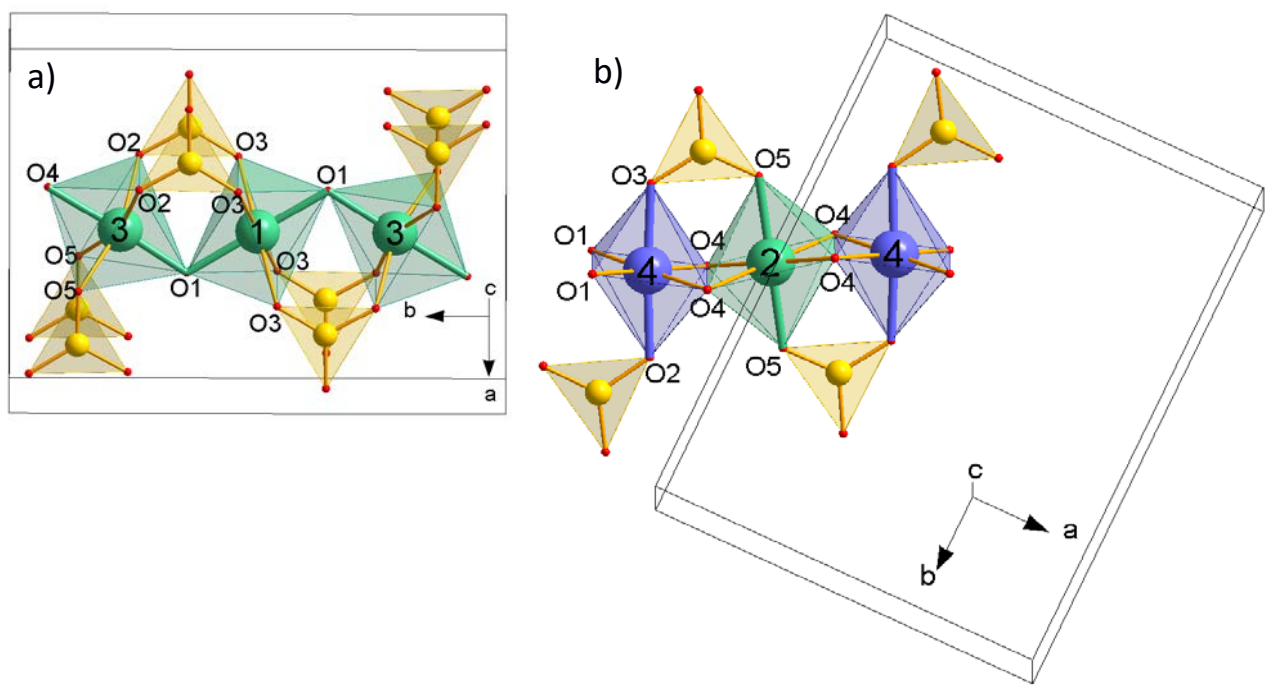

Fig. 13. The metal and boron coordination in Co$_3$BO$_5$ ludwigite. The triads Co3-Co1-Co3 (a) and Co4-Co2-Co4 (b) formed by divalent and trivalent ions are shown. The main octahedral axes along which the compression or elongation is observed are highlighted in bold.

Heating up the sample above 400 K triggers a stretching of equatorial Co3-O bonds and thus indirectly a compression of equatorial Co1-O ones inside the 3-1-3 triad. In addition there is an elongation of the axial Co1-O bonds. As a result, the Co1O$_6$ becomes more regular, while the Co3O$_6$ shows the opposite tendency. The other subsystem formed by Co$^{2+}$ and Co$^{3+}$ in sites 2 and 4 behaves in a different way. Above 400 K the local distortion of the Co4 site rises sharply while the $V_{zz}(2)$ for Co$^{2+}$ ions drops. As a result, the difference in the octahedral environments of two metal sites decreases and becomes negligible at 703 K. The thermal variations of the Co4O$_6$ bond-lengths are shown in Fig. 12c. It is clearly seen that the axial bonds, directed along BO$_3$ groups, grow much faster than equatorial ones resulting in the dramatic changes in $V_{zz}$.

To evaluate the oxidation states of different Co ions, a bond-valence-sum (BVS) approach has been applied [57]. The room-temperature oxidation states of Co1, Co2, and Co3 suggest that these sites are occupied by the Co$^{2+}$ ions, whereas the Co4 site is filled by Co$^{3+}$ (atomic charges 2.78(6)/2.84(7) when bond valence parameters are related to Co$^{2+}$/Co$^{3+}$ oxidation state) (Table 3). With temperature increases the valence bond charges for Co1, Co2, and Co3 sites tend to stay in the vicinity of +2, while the valence state of Co4 site drastically decreases (Fig. 12d). At $T$=703 K the empirical estimates predict atomic charges of 2.25(9)/2.30(7) for Co4. These values indicate propensity of site Co4 to shelter Co$^{2+}$ at high temperature phase. The empirical estimates of the valence for boron site are consistent with +3. The oxygen atomic charges show a clear decrement with temperature increase.

Table 3. Oxidation states of the metal and oxygen ions in Co$_3$BO$_5$ obtained using the BVS approach, $R_0$(Co$^{2+}$)=1.692 Å, $R_0$(Co$^{3+}$)=1.700 Å, $R_0$(B$^{3+}$)=1.371 Å, and $b$=0.37.

|     | 296 K | 703 K |
|-----|-------|-------|
|     | Co$^{2+}$/Co$^{3+}$ | Co$^{2+}$/Co$^{3+}$ |
| Co1 | 2.06(3)/2.10(8) | 2.05(3)/2.09(8) |
| Co2 | 2.27(8)/2.32(8) | 2.10(3)/2.14(9) |
| Co3 | 2.12(2)/2.16(9) | 2.09(0)/2.13(5) |
| Co4 | **2.78(6)/2.84(7)** | **2.25(9)/2.30(7)** |
| B   | 2.95(7) | 2.94(6) |
| O1  | 1.99(0) | 1.85(5) |
| O2  | 1.98(9) | 1.87(0) |
| O3  | 2.05(8) | 1.96(3) |
| O4  | 1.95(5) | 1.78(0) |
| O5  | 2.10(5) | 1.95(3) |

For the future discussion, note that we have observed negative thermal expansion of $a$-parameter in other homometallic oxyborate Mn$_2$BO$_4$ within the same $T$ interval (300-500 K) [58]. The warwickites and ludwigites have strong structural affinities and belong to the "3Å wall-paper structures". In both compounds the planar BO$_3$ triangles parallel to (110) plane connect adjacent layers of metal ions in [100] direction by corner-sharing. In ludwigites M4 sites located in the voids between the BO$_3$-tunnels are filled by M$^{3+}$ ions, whereas in warwickites similar sites are occupied by both di- and trivalent ions. From the crystal-chemical background it can be expected that the structure should be more rigid in the $a$-direction due to the rigidity of the bonds within the boron group and that the thermal/substitution effects should be more pronounced in $b$- and $c$- directions. This is confirmed by the comparison of values of thermal expansion coefficients for both materials. So, we can assume that the observed anomalies of *a*-parameters result from the structural peculiarities of the oxyborate family.

At the same time the thermal expansion coefficients of *b*-, *c*-, and unit cell have non-monotonic dependence on the temperature, going through a maximum at $T_1$=500 K where heat capacity and conductivity show the anomalies. This correlates with a rapid increase in the octahedral Co-O distances at Co4 site, occupied by $Co^{3+}$ and points out a manifestation of another mechanism of lattice expansion besides the conventional lattice anharmonicity. This mechanism can be attributed to the spin fluctuations arising from the change in the ionic radius of $Co^{3+}$ ion at spin-state transition from LS to HS. It is instructive to compare the LS-HS transition of the $Co^{3+}$ ion discussed here with simpler perovskite rare-earth cobaltites *Ln*CoO$_3$ (Ln=La, Dy, Sm, Gd, …) [59,60,61 and herein]. In LaCoO$_3$ a spin-state transition is well known and reveals itself with a sharp peak in the magnetic susceptibility at 150 K and smooth metallization above 500 K [62,63]. The maximum thermal expansion coefficient results from strong thermal fluctuations of the $Co^{3+}$ ions multiplicity between the low and high spin terms. All thermodynamic properties including the thermal expansion, magnetic susceptibility, specific heat are correlated with the spin-state transition in rare-earth cobaltites [64]. Moreover, maxima in the thermal expansion and magnetic susceptibility are related to the maximal rate of the multiplicity fluctuations when the derivative of the concentration of thermally excited HS terms $dn_{HS}/dT$ has maximum. This comparison indicates that in the $Co_3BO_5$ the maximal rate of the multiplicity fluctuations occurs at the temperature close to $T_1$=500 K. Upon further heating a clear anomaly is revealed in the temperature dependences of the expansion coefficients and the heat capacity with the maximum at $T_2$=700 K. Using the same methodology as for perovskite-like cobaltites, the observed anomaly can be associated with the smooth electronic transition caused by the modification of electronic structure. The conductivity data on $Co_3BO_5$ presented in Ref. [30] shows a gradual increase in σ(T) up to the highest measured temperature of 505 K. The electrical measurements data at elevated temperatures are not currently available.

The electron count in a triad 4-2-4 corresponds to three $Co^{3+}$ ions with one extra electron per triad. This itinerant electron can be localized at Co2 site, smeared over all three sites, or distributed between two adjacent Co2-Co4 sites leading to the dimerization state. In the $Fe_3BO_5$ a dimer formation at the 4-2-4 ladder is associated with structural phase transition and change in space group from *Pbam*(№55) to *Pbnm*(№62) at $T_{ST}$=283 K [9,10]. Neither present nor previous XRD studies reveal any structural phase transition over the entire investigated temperatures confirming the absence of the dimerization of the 4-2-4 triad in $Co_3BO_5$. Nevertheless, the spatial distribution for Co oxidation states in the system is dramatically changed leading to the electronic transition. The room-temperature BVS calculations clearly indicate the localized character of 3*d* electrons at Co2 and Co4 sites (the atomic charges are about +2.29 and +2.81, respectively) highlighting the charge ordering. Upon heating the oxidation state of Co4 site first decreases rapidly within *T* range 400-500 K and then more slowly, while the oxidation states of the rest of cobalt sites are almost unchanged. The increase in temperature was found to suppress the charge-ordered phase by means of squeezing out the excess charge from the Co4 octahedra. As a result, at 703 K all metal sites in the lattice become occupied by divalent ions, which inevitably should lead to the occurrence of some amount of holes at oxygen atoms. This effect manifests itself in the decrement of the oxygen atomic charges, most noticeable for O4 site. Now, Co2 and Co4 site are no longer distinguished but they become equivalent from the point of view of both octahedral distortions and atomic charges leading to the quenching of the charge ordering. This transition is not associated with the redox process, as one can see from DSC measurements, where the sample shows thermal stability over the measured temperature range, but it is more likely that it can be considered as an electronic transition caused by the reorganization of the electronic structure of $Co_3BO_5$. The electron transition found from X-ray diffraction involves not only metal sites but also oxygen ones in the lattice and should be seen from the heat-capacity, conductivity and spectroscopic measurements. The next finding is that the observed effective moment $\mu_{eff}$ = 4.87 $\mu_B$ extracted from high-temperature range T=500-700 K [30] can be reasonably explained. Indeed, if we assume that at high temperatures all cobalt ions become a divalent $n_{Co^{2+}}$=3 and have some orbital moment inherent in $Co^{2+}$ then $\mu_{eff}$ =4.84 $\mu_B/Co^{2+}$ (left panel of Fig. 8).

Within the ternary system Co-B-O, the following borate/oxyborates were obtained at ambient pressure $Co_2B_2O_5$ [47], $Co_3B_2O_6$ [48], $Co_3BO_5$ [4], $Co_4B_6O_{13}$ [49], α-$CoB_4O_7$ [65], and β-$CoB_4O_7$ [66] and HP-$CoB_2O_4$ [67] that require high pressure. Of these, only ludwigite contains trivalent cobalt ions. More interesting is that other 3*d* transition metals Fe, Mn, V, Cr, Ti form $M^{III}$-B-O systems isotypic calcite ($M^{III}BO_3$) [68-71], warwickite ($M^{II}M^{III}OBO_3$) [50,51], ludwigite ($M^{II}_2M^{III}O_2BO_3$) [3,8,72], and norbergite ($M^{III}_3BO_6$) [73,74,75]. This comparison indicates a clear tendency for cobalt systems to shelter $Co^{2+}$ ions. The thermal instability of $Co^{3+}$ in the borate systems requires additional study.

Another aspect that we would like to discuss is the effect of M4 ion on the long-range crystal structure. While the influence of M4 ion on the magnetic order in the ludwigites was discussed in Sections 4.1. and 4.2., here, we focus on the available structural data on the Co-based ludwigites. Taking the mean bond-length <Co-O> as a size (effective radius) of a given $CoO_6$ octahedron we have estimated the influence of the temperature and the cation size on the crystal structure. We checked the dependencies of the lattice parameters and cell volume on the effective size of metal ions located at different sites M1, M2, M3, and M4 and found that only the size of the M4 site causes a monotonic change in the crystal parameters (Fig. 14). The larger the effective size at M4 site, the larger the lattice parameters and volume. The substitution of smaller $Co^{3+}$ ion for larger $Ga^{3+}$, $Mn^{3+}$, $Fe^{3+}$, $Co^{2+}/Ti^{4+}$, $Co^{2+}/Sn^{4+}$ modifies the structural properties in the same way as the thermal expansion of pure $Co_3BO_5$ above 500 K. This conclusion is suitable for all Co-substituted ludwigites and clearly indicates the crucial role of the M4 ion in long-range crystal structure.

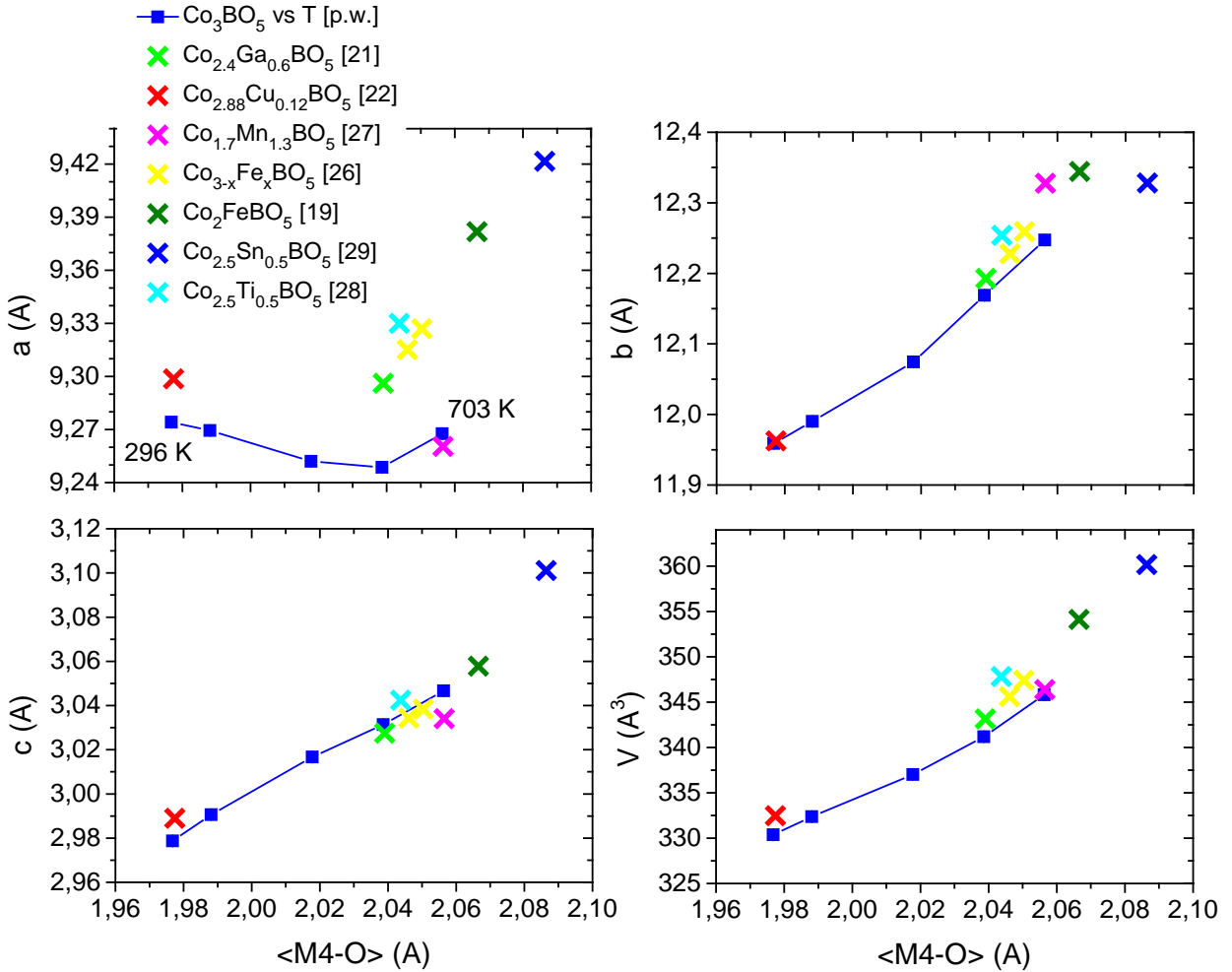

Fig. 14. The dependences of the lattice parameters and volume of Co-contained ludwigites on the M4 metal size.

### 4.5. Results of the DFT+U calculations

To understand the obtained results and determine the magnetic and electronic structure of $Co_3BO_5$, complementary theoretical studies have been carried out. The calculations were done using the GGA+U approach for both low- and high-temperature phases. For this the crystal structures corresponding to T=296 and 703 K, respectively, obtained in our X-ray diffraction experiments and magnetic structure determined in Ref. [29] were utilized.

Thanks to the ability of the GGA+U approach to stabilize different solutions corresponding not only to the global, but also different local minima of the density functional, we calculated total energies of two configurations, where Co4 ion is either magnetic or not (these configurations were obtained in conventional self-consistent DFT+U calculations without any constrains). These configurations are dubbed as ↓↑↑↓ or ↓↑↑0, where ↑ stands for spin up and ↓ for spin down and the order of signs corresponds to labels of the Co ions.

We found that the low temperature phase ↓↑↑0 configuration with nearly nonmagnetic Co4 ions has the lowest total energy, while ↓↑↑↓ solution is 46.88 meV per cell (12 Co ions) or 11.72 meV per Co4 ion higher in energy. Taking into account that the calculations were performed for the structure corresponding to $T$=296 K one may expect that this configuration can be thermally excited at temperatures ~432 K, i.e. close to the first anomaly in unit cell parameters observed experimentally.

Careful analysis of the occupation matrixes shows that there is a charge ordering in this low temperature phase: Co1, Co2, and Co3 are 2+ and have $3d^7$ electronic configuration, while Co4 is 3+ with six $3d$ electrons in both solutions. Moreover, Co4 ion is in the low-spin state in ↓↑↑0 and the high-spin state in ↓↑↑↓ configuration. List of magnetic moments on each Co ion can be found in Tab. 4. The density of states plots for both configurations are shown in Fig. 15. One may see that the system is insulating with the band gap of 1.4 eV for ↓↑↑0 configuration corresponding the ground state. This is rather important since the spin-state transition by itself does not lead to the metal-insulator transition. As we will show below only breaking of the charge ordering results in formation of a metallic state. The obtained value of the insulating gap is in reasonably good agreement with the experimental value of $E_g = 2 \cdot E_a \approx 1.7$ eV previously found from conductivity measurements below room temperature [76].

On another hand, we see that the charge ordering is not susceptible to the spin state of Co4 ion: in both calculated configurations this ion adopts 3+ charge state. Moreover, we performed additional calculations of other magnetic structures and obtained that even in fully ferromagnetic state Co4 ions are in $3d^6$ electronic configuration. Thus, we see that the mechanism of charge disproportionation is related with features of the crystal structure.

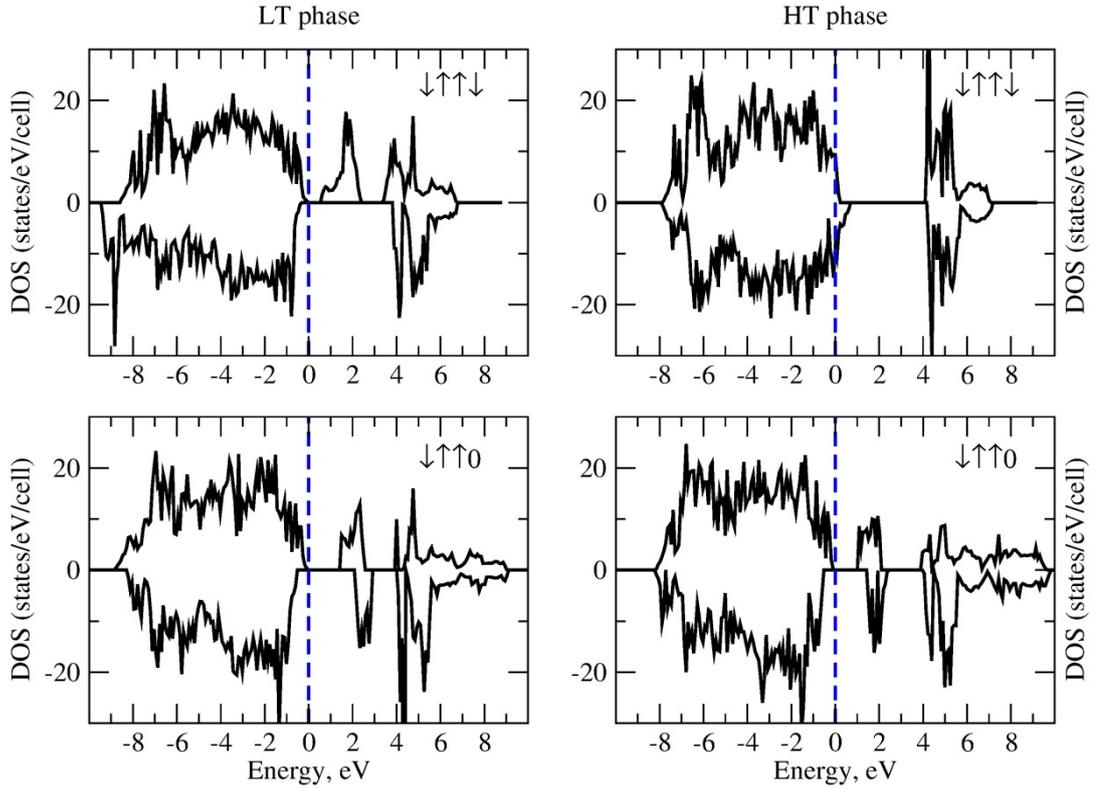

Fig. 15. $Co_3BO_5$ total DOS for LT and HT phases.

Table 4. Magnetic moment ($\mu_B$) of Co ions in $Co_3BO_5$ obtained for high- and low-temperature phases in the GGA+U calculations.

|     | R(L)T phase | | HT phase | |
| --- | --- | --- | --- | --- |
|     | ↓↑↑↓ | ↓↑↑0 | ↓↑↑↓ | ↓↑↑0 |
| Co1 | 2.8 | 2.8 | 2.8 | 2.8 |
| Co2 | 2.8 | 2.7 | 2.8 | 2.8 |
| Co3 | 2.8 | 2.8 | 2.8 | 2.8 |
| Co4 | 3.2 | 0.2 | 2.6 | 0.2 |

As it has been explained above local surroundings of the Co1 and Co3 are very different from Co2 and Co4. There are four $O^{2-}$ ions having bonds with $B^{3+}$ ions for Co1 and Co4 ions, but only two for Co2 and Co4. Then, effective electrostatic field acting on the Co1 and Co3 is weaker and already from these geometrical arguments one may expect that average Co-O distances for these ions should be longer than for Co2 and Co4 and, thus, Co1 and Co3 are prone to be 2+. This type of reasoning fully agrees with actual experimental Co-O bond distances, presented in Fig. 12. From the same Fig. 12 we see that this is Co4 ion, which has the smallest average Co-O bond length and may adopt 3+ charge state, which ionic radius (for any spin-state) is less than the one of $Co^{2+}$ ions.

However, with increase of the temperature the difference between average Co-O bond distances for different Co ions nearly vanishes destabilizing charge ordering. Our GGA+U calculations for the high temperature phase show that the charge ordering does disappear, all Co ions have $3d^7$ electronic configuration, and the system becomes metallic in the ↓↑↑↓ configuration (see Fig. 15), which now has the lowest total energy. Disappearance of the charge ordering may explain the experimentally observed increase of electrical conductivity [31]. It has to be mentioned that with given chemical formula one cannot have an insulating state with all Co ions being 2+, while in case of metal this is possible. But if such a state will be probed by, e.g., X-ray spectroscopy then there must be seen a lot of ligand holes, compensating 2+ valence state.

## 5. CONCLUSION

We have studied the electronic and magnetic state of single-crystalline $Co_3BO_5$ using X-ray magnetic circular dichroism, magnetic susceptibility, X-ray diffraction, differential scanning calorimetry, and heat capacity measurements. The experiments were done in a wide temperature interval covering both magnetically ordered and paramagnetic phases. The Co $K$-edge XANES/XMCD experiments were performed in a ferrimagnetic state. The extracted $XANES_{Co}^{3+}$ spectrum is an experimental evidence of the $Co^{3+}$ ions in $Co_3BO_5$. The Co sublattice magnetism has a similarity in two isostructural compounds $Co_3BO_5$ and $Co_2FeBO_5$, in the latter the $Co^{3+}$ ions are replaced by $Fe^{3+}$ ones. The rather small contribution from $Co^{3+}$ ion to XMCD signal of $Co_3BO_5$ is indicative of the LS state of this ion. The Co and Fe $K$-edges XMCD measurements recorded in the $Co_2FeBO_5$ single crystal have uncovered an antiferromagnetic coupling between the two sublattices, which indicates the generality of the formation of magnetic structures in $Co_3BO_5$ and $Co_2FeBO_5$, namely the ferrimagnetic arrangement of $Co^{2+}$ moments and antiferromagnetic arrangement of their moment relative to the magnetic moment at the M4 site. The sample-oriented measurements of magnetic susceptibility have revealed that a magnetic behavior of $Co_3BO_5$ in the paramagnetic regime (T<300 K) is dominated by the high uniaxial anisotropy. The effective magnetic moment 6.1 $\mu_B$/f.u. is compatible with two $Co^{2+}$ ions in the HS state and orbital contribution.

Our measurements of the thermodynamic properties of $Co_3BO_5$ have explored two anomalies at $T_1 \approx 500$ and $T_2 \approx 700$ K besides the magnetic phase transition at $T_N$. In contrast to work [30] the high-temperature anomalies we observed in $Co_3BO_5$ are smooth, which indicates that they are not first/second-order phase transitions. In line with this finding, our DSC study of the single crystal heated up to 770 K, confirmed that the sample is thermally stable. These anomalies detected by heat-capacity also manifested itself in the X-ray diffraction measurement, which reveals well-defined maxima of thermal expansive coefficients. We also observed a remarkable sensitivity of Co4 octahedral site to the temperature effect consisting in the change of the octahedral distortion and the oxidation state. With the increase in temperature the difference between Co2 and Co4 sites nearly vanishes making them equivalent. This is an electronic transition that is not caused by electron-sharing between the metal atoms in 4-2-4 triad as it was proposed in case of $Fe_3BO_5$ at

$T_{CO}$, but it is a consequence of the transition of the charge excess from Co4 metal site to the oxygen atoms (charge transfer). The available data do not allow determining whether the spin transition and the electronic transition occur simultaneously or they are spaced apart in temperature. The GGA+U calculations show that at the low-temperature phase the $Co_3BO_5$ is an insulator with the band gap of 1.4 eV, the ground state is ferrimagnetic with $Co^{3+}$ ion is being in LS state. At high temperature the charge ordering does disappear, all Co ions have $3d^7$ electronic configuration, and the system becomes metallic.

### ACKNOWLEDGMENTS

Natalia Ivanova's and Leonard Bezmaternykh's memory is dedicated.
The authors acknowledge A. Ney for the allow us to use XANES spectrum of $ZnCo_2O_4$ film. We are grateful to the Russian Foundation for Basic Research (project no. 20-02-00559), and President Council on Grants (project no. MK-2339.2020.2) for supporting our work. This research was carried out within the state assignment of the Russian Ministry of Science and High Education via program "Quantum" (No. AAAA-A18-118020190095-4). We also acknowledge support by Russian Ministry of Science via contract 02.A03.21.0006. We acknowledge financial support from the Spanish MINECO DWARFS project MAT2017-83468-R, and Gobierno de Aragón (RASMIA group, E12-17R).